\documentclass[11pt,pdftex,a4paper]{article}

\usepackage[margin=1in]{geometry}
\usepackage{epic,eepic}
\usepackage{graphics}
\usepackage{graphicx}
\usepackage{multicol}
\usepackage{xcolor}
\usepackage{subcaption}
\usepackage{hyperref}	
\usepackage{amsmath}
\usepackage[vlined,ruled,linesnumbered,noresetcount]{algorithm2e}
\usepackage{cleveref}
\usepackage[normalem]{ulem}
\usepackage{listings}
\usepackage{subcaption}

\newcommand{\remove}[1]{}

\newcommand{\lred}[1]{{\color{red} #1}\normalcolor}
\newcommand{\lblue}[1]{{\color{blue} #1}\normalcolor}

\newcommand{\lblack}[1]{{\color{black} #1}\normalcolor}

\newtheorem{definition}{Definition}
\newcommand{\qedsymb}{\hfill{\rule{2mm}{2mm}}}

\SetKwProg{myproc}{Procedure}{}{}

\DontPrintSemicolon

\let\oldnl\nl
\newcommand{\nonl}{\renewcommand{\nl}{\let\nl\oldnl}}

\makeatletter
\newcommand{\removelatexerror}{\let\@latex@error\@gobble}
\makeatother

\makeatletter
\lst@Key{countblanklines}{true}[t]%
{\lstKV@SetIf{#1}\lst@ifcountblanklines}

\lst@AddToHook{OnEmptyLine}{%
	\lst@ifnumberblanklines\else%
	\lst@ifcountblanklines\else%
	\advance\c@lstnumber-\@ne\relax%
	\fi%
	\fi}
\makeatother

\newcommand{\pwb}[1]{\textbf{pwb(#1)}}
\newcommand{\PWB}[1]{pwb}
\newcommand{\pfence}[1]{\textbf{pfence(#1)}}
\newcommand{\psync}[1]{\textbf{psync(#1)}}
\newcommand{\pwbi}{\mbox{\texttt{pwb}}}
\newcommand{\pfencei}{\mbox{\texttt{pfence}}}
\newcommand{\psynci}{\mbox{\texttt{psync}}}

\newcommand{\CAS}{\mbox{\textit{CAS}}}

\newcommand{\LL}{\mbox{\textit{LL}}}
\newcommand{\SC}{\mbox{\textit{SC}}}
\newcommand{\LLSC}{\mbox{\textit{LL/SC}}}
\newcommand{\LLVLSC}{\mbox{\textit{LL/VL/SC}}}
\newcommand{\VL}{\mbox{\textit{VL}}}
\newcommand{\True}{\mbox{\texttt{true}}}
\newcommand{\False}{\mbox{\texttt{false}}}

\newcommand{\PWFcomb}{{\sc PWFcomb}}
\newcommand{\PWFqueue}{{\sc PWFqueue}}
\newcommand{\PWFstack}{{\sc PWFstack}}
\newcommand{\PWFenqueue}{{\sc PWFenqueue}}
\newcommand{\PWFdequeue}{{\sc PWFdequeue}}
\newcommand{\EnqueueConnect}{{\sc EnqueueConnect}}
\newcommand{\DequeueConnect}{{\sc DequeueConnect}}

\newcommand{\PBcomb}{{\sc PBcomb}}
\newcommand{\PBqueue}{{\sc PBqueue}}
\newcommand{\PBheap}{{\sc PBheap}}
\newcommand{\PBstack}{{\sc PBstack}}
\newcommand{\PBqueueEnq}{{\sc PBqueueEnq}}
\newcommand{\PBqueueDeq}{{\sc PBqueueDeq}}
\newcommand{\PSim}{{\sc PSim}}
\newcommand{\SimQueue}{{\sc SimQueue}}
\newcommand{\HSynch}{{\sc H-Synch}}
\newcommand{\CCSynch}{{\sc CC-Synch}}

\newcommand{\MSQueue}{{\sc MSQueue}}
\newcommand{\OneFile}{{\sc OneFile}}
\newcommand{\CXPUC}{{\sc CX-PUC}}
\newcommand{\CXPTM}{{\sc CX-PTM}}
\newcommand{\RedoOpt}{{\sc RedoOpt}}
\newcommand{\FHMP}{{\sc FHMP}}
\newcommand{\NormOpt}{{\sc NormOpt}}
\newcommand{\CAPSULES}{{\sc Capsules-Normal}}
\newcommand{\Romulus}{{\sc Romulus}}
\newcommand{\DFC}{{\sc DFC}}
\newcommand{\Bcomb}{{\sc Bcomb}}

\newcommand{\LDQ}{{\sc OptLinkedQ}}
\newcommand{\UDQ}{{\sc OptUnlinkedQ}}
\newcommand{\PBstackNE}{{\sc PBstack-no-elim}}
\newcommand{\PWFstackNE}{{\sc PWFstack-no-elim}}
\newcommand{\PBstackNR}{{\sc PBstack-no-rec}}
\newcommand{\PWFstackNR}{{\sc PWFstack-no-rec}}

\newcommand{\Recover}{{\sc Recover}}

\newcommand{\Mark}{\mbox{\sc Mark}}
\newcommand{\UnMark}{\mbox{\sc UnMark}}
\newcommand{\IsMarked}{\mbox{\sc IsMarked}}
\newcommand{\AtomicFloat}{\mbox{\tt AtomicFloat}}

\SetKw{Continue}{continue}
\SetKw{Break}{break}
\SetKw{WaitUntil}{wait until}
\SetKw{And}{and}
\SetKw{Or}{or}
\SetKw{Mod}{mod}
\SetKw{Allocate}{allocate}
\SetKw{Execute}{execute}
\SetKw{Using}{using}
\SetKw{Replacing}{replacing}
\SetKw{Adding}{adding}
\SetKw{Add}{add}
\SetKw{To}{to}
\SetKw{CASS}{CAS}

\newcommand{\comnospace}{\mbox{$\triangleright$}}
\newcommand{\com}{\mbox{\comnospace\ }}

\newcommand{\Enqueue}{\mbox{\sc Enqueue}}
\newcommand{\Dequeue}{\mbox{\sc Dequeue}}
\newcommand{\Push}{\mbox{\sc Push}}
\newcommand{\Pop}{\mbox{\sc Pop}}
\newcommand{\HInsert}{\mbox{\sc HInsert}}
\newcommand{\HDelete}{\mbox{\sc HDeleteMin}}
\newcommand{\HGet}{\mbox{\sc HGetMin}}

\newcommand{\PerformRequest}{\mbox{\sc PerformRequest}}
\newcommand{\PerformEnqueueRequest}{\mbox{\sc PerformEnqReq}}
\newcommand{\PerformDequeueRequest}{\mbox{\sc PerformDeqReq}}

\newcommand{\Type}{\textbf{type}}

\title{Persistent Software Combining}

\date{}

\author{
	Panagiota Fatourou\\
	Universit\'e Paris Cit\'e, LIPADE, France\\
	FORTH ICS and University of Crete, Greece\\
	faturu@csd.uoc.gr
	\and
	Nikolaos D. Kallimanis\\
	FORTH ICS\\
	nkallima@ics.forth.gr
	\and
	Eleftherios Kosmas\\
	University of Crete, Greece\\
	ekosmas@csd.uoc.gr
}

\begin{document}

\maketitle
\begin{abstract}
The availability of Non-Volatile Main Memory (known as NVMM) 
enables the design  of recoverable concurrent algorithms. 
We study the power of 
software combining in achieving recoverable synchronization 
and designing persistent data structures.
{\em Software combining} is a general synchronization approach,
which attempts to simulate the ideal world when executing
{\em synchronization requests} (i.e., requests that must be executed
in mutual exclusion). A single thread, called the {\em combiner}, 
executes all active requests, while the rest of the threads are waiting
for the combiner to notify them that their requests have been applied. 
{\em Software combining} significantly decreases the synchronization
cost and outperforms many other synchronization techniques
in various cases.  

We identify three persistence principles, crucial for performance, 
that an algorithm's designer has to take into consideration
when designing highly-efficient recoverable synchronization protocols or data structures.  
We illustrate how to make the appropriate design decisions in all stages of
devising recoverable combining protocols to respect these principles.
Specifically, we present two recoverable software combining protocols, satisfying 
different progress properties, 
that are many times faster and have much lower persistence cost than a large collection 
of existing persistent techniques for achieving scalable synchronization. 
We build fundamental recoverable data structures, such as stacks and queues,
based on these protocols that outperform {\em by far} existing recoverable implementations of such data structures.
We also provide the first recoverable implementation of a concurrent heap and 
present experiments to show that it has good performance when the size of the heap
is not very large.
\end{abstract}

\section{Introduction}
\label{sec:intro}

Recent advances in memory technology have resulted in 
byte-addressable Non-Volatile Main Memory (NVMM), which
attempts to combine the performance benefits of conventional 
main memory with the strong persistence characteristics of secondary storage. 
A program running in a traditional memory hierarchy system stores its operational data 
in volatile data structures maintained in DRAM, whereas its recovery data (such as transactional logs) 
are usually stored in non-volatile secondary storage. 
In the event of a failure, all in-memory data structures are lost 
and must be re-constructed from recovery data to make the system functional again. 
This poses major performance overheads.
The availability of NVMM enables the design of
concurrent algorithms, 
whose execution will be recoverable at no significant cost. 
An algorithm is  {\em recoverable} (also known as {\em persistent}~\cite{CC+11-I}
or {\em durable}~\cite{VenkataramanTRC-FAST2011}) if its state can be restored after 
recovery from a system-crash failure. 
Another important property, known as {\em detectability}~\cite{AttiyaBH-PODC2018,FriedmanQueue18,LG21}, 
is to be able to determine, upon recovery, if an operation has been completed, 
and if yes, to find its response. 
Despite many efforts for designing efficient recoverable
synchronization protocols and data structures (see Section~\ref{sec:rel}), persistence comes
at a significant cost even for fundamental data structures, such as 
stacks and queues. 

When designing recoverable algorithms, the main challenge stems from 
the fact that data stored into registers and caches are volatile.
Thus, unless they have been flushed to persistent memory, 
such data will be lost at a system crash. Flushing to persistent memory
occurs by including specific {\em persistence instructions}, such as 
\pwbi, \pfencei\ and \psynci\ in the code, which are however expensive in terms of performance.

In this paper, we reveal the power of 
software combining in achieving recoverable synchronization 
and designing persistent data structures.
In {\em software combining}~\cite{FK12ppopp,HIST10,OTY99,FK17opodis,KSW18},
each thread first announces its request, and then tries to become the combiner
by acquiring a lock. 
The combiner applies several active requests, in addition to its own, before it
releases the lock.
As long as the combiner serves active requests, other threads
perform local spinning, waiting for the combiner to release the lock. 
As soon as the lock is released, waiting threads whose requests 
have been served  by the combiner, return the calculated responses,
whereas the rest 
compete again for the lock. 
Software combining~\cite{FK12ppopp,FK17opodis} has been 
proved to outperform many other synchronization techniques
in various cases, and  
has been used to implement state-of-the-art fundamental concurrent data structures,
such as queues and stacks~\cite{FK12ppopp,FK17opodis},
that lie in the heart of inter-thread communication mechanisms.

Although simple in their nature, combining protocols
should be designed carefully, as they encompass five design decisions that 
all may have crucial impact in performance. 
Existing combining protocols differ in these design decisions,
exhibiting different performance~\cite{FK12ppopp,HIST10,KSW18,OTY99}. 

\begin{definition}
\label{dd}
Design decisions for combining protocols that are crucial for performance:
\begin{enumerate}
\item the mechanism to decide which of the active threads 
will act as the combiner (e.g., some combining protocols use \CAS~\cite{HIST10,FK11spaa,FK14}, others use queue locks~\cite{FK12ppopp}); 
\item the data structure to store the active requests; 
\item how the updates are applied (e.g., directly on the shared state or on a copy of it); 
\item the mechanism for collecting the requests' responses;
\item how to discover which requests have not been applied. 
\end{enumerate}
\end{definition}

In this paper, we present two {\em recoverable} software combining protocols,
\PBcomb\ which is blocking, and \PWFcomb\ which is wait-free. 
We designed all five stages of our protocols taking into consideration three principles 
for reducing persistence cost (motivated
by our experiments; also discussed in~\cite{SP21,KJI+21,YK+20,AB+19,AB+new}),
that are presented in Definition~\ref{pp}. 
Our experiments show that the resulting protocols are many times faster
than a large collection of existing persistent techniques for achieving scalable synchronization. 
\begin{definition}
\label{pp}
Persistence principles crucial for performance:
\begin{enumerate}
\item The number of the persistence instructions 
should be maintained as low as possible. This encompasses that
an implementation must store in NVMM only those variables 
(and persist those values of them) that are necessary for recoverability. 
\label{low number}
\item  The persistence instructions should be of low cost. Not all persistence 
instructions have the same cost~\cite{AB+new,SP21,YK+20}. For instance, reducing 
contention on non-volatile variables can be beneficial for performance~\cite{AB+new,SP21}. \label{low cost}
\item Data to be persisted should be placed in consecutive 
memory addresses, so that they are persisted all together~\cite{KJI+21}. 
\label{consecutive}
\end{enumerate}
\end{definition}
\vspace*{-.5cm}
Combining is a promising approach for achieving persistent
synchronization at low cost, as having no more than the combiner thread persisting
updates on the state of the implemented object is expected to reduce the number of 
persistence instructions that are performed, as well as to decrease contention on
persisted data.
However, the design decisions of state-of-the-art combining protocols~\cite{FK12ppopp,HIST10,FK17opodis,KSW18}
are not fully in favor of supporting persistence in an efficient way: 
All these protocols store the active requests in a dynamic linked list, and have the combiner traversing the list 
to figure out which requests are active. Moreover, the combiner applies the active requests 
on the shared state of the object, and records responses in the 
list nodes. 
When attempting to make these protocols recoverable without changing their design decisions,
the updated shared state, and the requests' responses that the combiners calculate
need to be persisted for ensuring recoverability. These data are scattered in memory.
This violates persistence principles~\ref{low number} and~\ref{consecutive},
introduces several complications that the designer needs to cope with (see e.g., ~\cite{DFC}),
and results in high persistence overhead (see Section~\ref{sec:perf}).

Our algorithms differ from existing state-of-the-art combining protocols
(including the \CCSynch~\cite{FK12ppopp} algorithm and flat-combining~\cite{HIST10}), 
illustrating how all five design decisions should take into consideration the three persistence
principles of Definition~\ref{pp}. This results in protocols that have low persistence cost,
in addition to being highly efficient in terms of synchronization.
Our experiments show 
that both, \PBcomb\ and \PWFcomb, outperform by far, many previous recoverable Transactional Memory (TM) Systems~\cite{CFR18,CFP20eurosys,PMDK-web,RC+19}
and several generic mechanisms for designing recoverable data structures~\cite{AB+19,AB+20,NormOptQueue19,SP21} proposed in the literature.
Specifically, \PBcomb\ is $4$x faster and \PWFcomb\ is 2.4x faster than the competitors.
Our protocols satisfy {\em detectable recoverability}~\cite{FriedmanQueue18}, 
whereas most competitors (all but~\cite{NormOptQueue19,AB+20})
guarantee only weaker consistency properties, such as {\em durable linearizability}~\cite{IMS16}.

We build recoverable queues and stacks using \PBcomb\ and  \PWFcomb.
Our experiments illustrate that the recoverable queues (\PBqueue\ and \PWFqueue) 
and stacks (\PBstack\ and \PWFstack) 
that are built on top of \PBcomb\ and \PWFcomb, have much better performance than  
state-of-the-art recoverable implementations of such data structures,
including the specialized recoverable queue implementations in~\cite{FriedmanQueue18,SP21}.
Concurrent queues and stacks play a significant role in runtime systems~\cite{AKD2012}, high performance computing~\cite{OpenMPI,OpenMPI-CCSynch}, 
kernel schedulers, network interfaces~\cite{PKA2019}, etc.
The proliferation of NVMM and the availability of highly-efficient recoverable stacks and queues
could enable persistence in such settings.

Based on \PBcomb, we were able to design the first recoverable concurrent heap (\PBheap);
experiments show that \PBheap\ has good performance when the heap is not too large.
\PBheap\ is useful for implementing recoverable versions of algorithms that rely on priority queues
when the problem input size is small or medium. 
Implementations of concurrent heaps
often do not scale well due to contention (mainly at the root node). This makes a heap implementation 
a natural candidate for applying software combining.

Our contributions are summarized as follows.
\begin{itemize}
\vspace*{-.1cm}
\item We present two highly-efficient recoverable combining protocols,
which exhibit low persistence overhead and small synchronization cost.

\item Experiments show that our protocols outperform {\em by far}
state-of-the-art recoverable universal constructions and software transactional systems
(that often ensure weaker consistency properties
than our algorithms).

\item 
We illustrate how to make the appropriate design decisions in all stages of
designing combining protocols to respect the three persistence principles, crucial for 
performance.
Our experiments reveal the performance power of respecting these principles. 

\item We built recoverable queues and stacks, based on 
our combining protocols, which outperform {\em by far} previous recoverable implementations
of stacks and queues, including specialized recoverable implementations
of such data structures~\cite{FriedmanQueue18,DFC,SP21}.

\item We provide the first recoverable implementation of a concurrent heap
and present experiments to show that, for small/medium heap sizes, it has good performance. 
\end{itemize}

\vspace*{-.2cm}
\section{Preliminaries}
\label{sec:prelim}

We consider a standard asynchronous distributed system with $n$ threads.
The system supports the atomic execution of {\em base primitives}, such as reads, writes, \CAS, 
and \LLVLSC\
on single-word shared variables. 
A \CAS($\mathit{O, old, new}$)\ checks if the state of object $O$ is equal to $\mathit{old}$
and if so, it changes it to $\mathit{new}$ and returns \True, otherwise the state
of $O$ remains unchanged and \False\ is returned. 
An \LLSC\ object $\mathit{O}$ supports the operations \LL\
(which returns the current value of $\mathit{O}$) and \SC. By executing $\SC(O,v)$, 
a thread $\mathit{p}$ attempts to set the value of $\mathit{O}$ to $v$. This change takes place only 
if no thread has changed the value of $\mathit{O}$ (by executing \SC) since the execution 
of $\mathit{p}$'s latest \LL\ on it; then, the \SC\ is successful and returns \True.
Otherwise the \SC\ returns \False.
We assume the Total Store Order (TSO) model, 
supported by x86 and SPARC, where writes by the same thread become visible in program order.

Current architectures supporting non-volatile main memory 
(e.g., those supporting Intel Optane DC Persistent Memory) 
provide both DRAM and NVMM.
System-wide crash failures may occur at any point in time. When a failure 
occurs, the values of all variables stored 
in volatile memory (e.g., in registers, caches, or DRAM)
are lost 
(upon recovery, these variables have their initial values),
whereas values that have been written back (or {\em persisted}) to NVMM are non-volatile.
Storing data in DRAM is desirable for good performance (Persistence Principle~\ref{low number}).

We assume \emph{explicit epoch persistency}~\cite{IMS16}:
a write-back to persistent memory is triggered by a persistent
write-back (\pwbi) instruction.
The order of
\pwbi s\ is not necessarily preserved. When ordering is required, a \pfencei\ instruction
can be used to order preceding \pwbi\ instructions before all subsequent \pwbi s. A thread executing a \psynci\ instruction
blocks until all previous \pwbi\ instructions complete.
For each shared variable, \pwbi s preserve program order. 
We call \pwbi, \pfencei, and \psynci, the {\em persistence instructions}.

Failed threads can be recovered by the system in an asynchronous way. 
A {\em recoverable} (or {\em persistent}) implementation provides, 
for each thread and for each supported operation $\mathit{op}$, an associated
\emph{recovery function}. Upon recovery, $\mathit{op}$'s recovery function is invoked by the system 
for each  thread that was executing an instance of $\mathit{op}$ at the time the system crashed. 
If a crash occurs while the recovery function of $\mathit{op}$ is executed,
the recovery function of $\mathit{op}$ is re-invoked.

An execution is {\em durably linearizable}, if the effects of all operations
that have completed before a system crash
are reflected in the object's state upon recovery (see~\cite{IMS16} for a formal definition). 
\emph{Detectability}~\cite{AttiyaBH-PODC2018,FriedmanQueue18,LG21}
ensures that it is possible to determine, upon recovery, 
whether an operation took effect, and its response value, if it did.
{\em Detectable recoverability} ensures durable linearizability and detectability.

Detectable recoverability cannot be achieved without system support~\cite{BHR20}. 
As in~\cite{BHR20, AB+new, DFC}, we assume that the system 
persists the information that is needed for calling, for every thread $\mathit{p}$, 
the recovery function for $p$ with the same arguments as the instance of $\mathit{op}$
that $\mathit{p}$ was executing at crash time.
Moreover, for compatibility with previous work~\cite{FriedmanQueue18}
(and fair treatment of the algorithms in the experimental analysis), 
we assume that each  thread $\mathit{p}$ has an associated persistent sequence number $\mathit{seq}$
which it increments each time it invokes an operation  $\mathit{op}$ and passes it as a parameter
to $\mathit{op}$.
The system invokes the recovery function for $\mathit{op}$ passing the same value for $\mathit{seq}$ as 
in the original invocation of $\mathit{op}$ by $\mathit{p}$. We remark that our algorithms also work
with just passing to each operation of $\mathit{p}$ a toggle bit (instead of $\mathit{seq}$) whose values 
alternate from one invocation of the thread to the next (i.e., just using the value
of the last bit of $\mathit{seq}$). Our algorithms can be adjusted to work also with 
other assumptions for system support that have been made in previous work~\cite{AB+19,BHR20}
for designing detectable implementations (see Section~\ref{sec:rel} for more details). 
Without any system support, our algorithms ensure durable linearizability (but not detectability).

A recoverable implementation is \emph{lock-free}, 
if in every infinite execution produced by the implementation, 
which contains a finite number of system crashes, 
an infinite number of operations complete. 
An execution is {\em wait-free}, if every operation 
completes within a finite number of steps if it does not experience any crash after 
some point of its execution.

\section{Blocking Combining and Recoverability}
\label{sec:PBcomb}

\noindent
{\bf Overview of \PBcomb.}
\PBcomb\ follows the general idea of blocking software combining~\cite{FK12ppopp,HIST10,OTY99,FK17opodis,KSW18}. 
\PBcomb\ achieves low synchronization cost, while respecting all persistence principles:
\begin{enumerate}

\item \PBcomb\ implements the lock in volatile memory.
We have chosen a lock implementation 
which aims mainly at reducing synchronization cost. 
Moreover, the lock implementation allows 
a thread to leave the entry-section without ever acquiring the lock,
if it finds out that its request has been served by a combiner.

\item \PBcomb\ utilizes an array, $\mathit{Request}$, to store the threads' 
requests in consecutive memory addresses. 
This array is stored in volatile memory (i.e., it does not have to be persisted). 
This results in lower persistence cost.

\item 
Each combiner creates a copy of the state 
of the implemented object and applies the active requests on this copy (and not on the shared state of the object).
This is one of the most crucial design decisions of \PBcomb\ in terms of performance. 
The combiner switches a shared variable to index the copy it used, 
indicating that it stores the current valid state of the implemented object.
The combiner should persist the copy it used
before trying to switch the pointer. 

There is an interesting performance tradeoff between the approach of performing updates directly on the shared state
and that of creating a copy of the state to apply the updates on.
In the first technique, the updates are performed on 
data that are usually scattered in memory. Persisting the updated values is thus expensive. 
This problem is avoided by the 
second technique which persists data stored in the copy in consecutive
memory addresses. However, the second technique works well mainly for 
objects of small or medium size (or when the number of synchronization points
is small). 
In other cases, the cost of copying and persisting the state 
may dominate the cost of persisting a smaller amount of scattered data (part of the state). 

A well-known limitation~\cite{FK11spaa,FK12ppopp,FK17opodis} of the combining technique is that 
using a single thread to apply all active requests
may restrict parallelism, if the size of the object or the number of synchronization points
are large. 
\PBcomb\ (similarly to previous persistent algorithms~\cite{DFC} that are based on some combining protocol), inherits 
the limitations of the technique. Thus, \PBcomb\ works well mainly for implementing objects of small and medium size
or when the number of synchronization points is small (as is the case with stacks and queues). 
Subsequently, creating a copy of the state significantly reduces the persistence cost
without imposing any additional limitation to the algorithm.

\item Following persistence principle~\ref{consecutive},
\PBcomb\ stores the response values in an array, maintained 
together with the state of the object (in consecutive memory addresses). 
The combiner persists the entire array of return values
together with the object's state. 

\item \PBcomb\ uses two bits ($\mathit{activate}$ and $\mathit{deactivate}$) for each thread $\mathit{p}$,  
to identify whether the last  request, initiated by $\mathit{p}$, has been served.
If the two bits are not equal, 
$\mathit{p}$ has a request which has not yet been served\, i.e., it is {\em active}.
\PBcomb\ persists just the deactivate bit of $\mathit{p}$. 
Following persistence principle~\ref{consecutive}, 
\PBcomb\ stores the deactivate bits together with the object's state, 
so all data to be persisted are in consecutive memory locations. 
\end{enumerate}

\vspace*{-.2cm}
We define the {\em combining degree}, $\mathit{d}$, to be the average number
of requests that a combiner serves.
\PBcomb\ executes a small number of \pwbi\ instructions
for every $\mathit{d}$ requests.
Moreover, in \PBcomb, threads other than the combiner do not have to execute any persistence instructions.
Additionally, the combiner does not persist each of the requests it applies separately;
data to be persisted are stored in consecutive memory addresses and are persisted all together.
Thus, \PBcomb\ respects the persistence principles, maintaining persistence cost low.
Additionally, \PBcomb\ has significantly lower synchronization cost than previous combining protocols~\cite{FK11spaa,FK12ppopp,HIST10},
as well as than its competitors; this is another major reason for its good performance. 

\begin{algorithm}[t]
\setcounter{AlgoLine}{0}
	\removelatexerror
	\scriptsize
	\begin{flushleft}
		\Type\ RequestRec \{ \label{alg:pbcomb-vol-rec:AnnounceRec-start}\;
		\hspace*{3mm} Function $func$\;
		\hspace*{3mm} Argument $Args$\;
		\hspace*{3mm} Bit $activate$\;
		\hspace*{3mm} Bit $valid$\;		
\nonl	\}							\label{alg:pbcomb:AnnounceRec-end}\;

		\vspace*{1mm}
		\Type\ StateRec \{\;
		\hspace*{3mm} State $st$\;
		\hspace*{3mm} ReturnValue $ReturnVal[0..n-1]$\;
		\hspace*{3mm} Bit $Deactivate[0..n-1]$\;
\nonl	\}\;
\nonl	\vspace*{1mm}
		\com Shared non-volatile variables:\;
		StateRec $MemState[0..1]$, initially $\langle\bot, \langle\bot,\ldots,\bot\rangle, \langle0,\ldots,0\rangle\rangle$\;
		Bit $MIndex$, initially $0$\; \nonl
		\vspace*{1mm}
		\com Shared volatile variable:\;
		RequestRec $Request[0..n-1]$,	initially $\langle\langle\bot,\bot,0,0\rangle, \ldots, \langle\bot,\bot,0,0\rangle\rangle$\;	\label{alg:pbcomb:announce-array}
		Integer $Lock$, initially $0$ \;
		Integer $LockVal$, initially $0$
	\end{flushleft}	

	\vspace*{2mm}


	 \myproc{ReturnValue {\footnotesize \PBcomb}(Function $func$, Argument $Args$, Integer $seq$)}{
		\tcp{Announce request}
		$Request[p] := \langle func, Args, 1-Request[p].activate,1 \rangle$ \;\label{alg:pbcomb:announce}
		\KwRet\ \PerformRequest()																\label{alg:pbcomb:response}
	}

	\vspace*{2mm}


	\myproc{ReturnValue {\footnotesize \Recover}(Function $func$, Argument $Args$, Integer $seq$)}{
\nl		$Request[p]:= \langle func, args, seq \mod 2, 1\rangle$\;					\label{alg:pbcomb:recovery:update-Request}
		\tcp{if request is not yet applied}
		\uIf {$MemState[MIndex].Deactivate[p] \neq seq \mod 2$} {					\label{alg:pbcomb:recovery:rec-not-applied}
			\KwRet\ \PerformRequest()							\label{alg:pbcomb:recovery:re-execute-rec}
		}
		\KwRet $MemState[MIndex].ReturnVal[p]$ \ \ \tcp{request is applied}				\label{alg:pbcomb:recovery:rec-applied}
	}
\caption{\PBcomb\ -- Code for thread $p \in \{0, \ldots, n-1\}$}
\label{alg:pbcomb-types}
\end{algorithm}

\begin{algorithm}[t]
\setcounter{AlgoLine}{0}
	\removelatexerror
	\scriptsize

	
	\myproc{ReturnValue {\footnotesize \PerformRequest()}}{
\nl		Bit $ind$   \tcp*{local variable of $p$} 								\label{alg:pbcomb:first_line}
		Integer $lval$ \tcp*{local variable of $p$}
		\While{\True} {																	\label{alg:pbcomb:while}
			$lval := Lock$ \;
			\uIf {$lval$ \Mod $2 = 0$} {								\label{alg:pbcomb:ckeck_even}
				\lIf {$CAS(Lock, lval, lval+1) = \True$} { \Break }				\label{alg:pbcomb:acquire_lock}
				$lval := lval+1$
			}

			\WaitUntil $Lock \neq lval$ \;							\label{alg:pbcomb:busy_wait}
			\uIf {$Request[p].activate = MemState[MIndex].Deactivate[p]$} {			\label{alg:pbcomb:if_req_processed}
				\lIf {$LockVal \neq lval$} {
					\WaitUntil $Lock \neq lval+2$
				}
				\KwRet $Memstate[MIndex].ReturnVal[p]$;					\label{alg:pbcomb:return}
			}
		}
	
		$ind: = 1 - MIndex$ \;							\label{alg:pbcomb:combiner:flip_index}
		$MemState[ind] := MemState[MIndex]$          \tcp*{copy current state}	\label{alg:pbcomb:combiner:state_copy}
		\For{$q\gets0$ \KwTo $n-1$} {								\label{alg:pbcomb:combiner:for:start}
			\tcp{if $q$ has a request that is not yet applied}
			\uIf {$Request[q].valid=1$ and $Request[q].activate \neq MemState[ind].Deactivate[q]$} {        		\label{alg:pbcomb:combiner:if_apply}
				{\bf apply} $Request[q].func$ with $Request[q].Args$ on $MemState[ind].st$ \label{alg:pbcomb:apply} \;
				{\bf compute} return value, $returnVal$ \;
				$MemState[ind].ReturnVal[q] := returnVal$ \;
				$MemState[ind].Deactivate[q] := Request[q].activate$ \;			\label{alg:pbcomb:deactivate}
			}
		}																				\label{alg:pbcomb:combiner:for:end}
		
		\pwb{$\&MemState[ind]$}				\label{alg:pbcomb:persist_MemState} \;
		\pfence{} \; 	\label{alg:pbcomb:sync_MemState}	
		$LockVal:=Lock$\;
		$MIndex := ind$ \;				\label{alg:pbcomb:combiner:updateS}
		\pwb{$\&MIndex$}				\label{alg:pbcomb:persist_MIndex}  \;		
		\psync{}					\label{alg:pbcomb:sync_MIndex} \;
	
		$Lock := Lock + 1$ \;				\label{alg:pbcomb:combiner:unlock}
		\KwRet $MemState[MIndex].ReturnVal[p]$													\label{alg:pbcomb:combiner:return}
	}
\caption{\PBcomb\ -- Code for thread $p \in \{0, \ldots, n-1\}$}
\label{alg:pbcomb-funcs-2}
\end{algorithm}

\noindent
{\bf Detailed Description.} \PBcomb\  appears in Algorithms~\ref{alg:pbcomb-types} and~\ref{alg:pbcomb-funcs-2}.
Each element of $\mathit{Request}$  stores a $\mathit{RequestRec}$ record with fields: 
i) a pointer $\mathit{func}$ to a function to execute in order to serve the request,
ii) a set $\mathit{args}$ of arguments to $\mathit{func}$,
iii) a bit $\mathit{activate}$ used to identify whether the request 
has already been served or not, and iv) a $\mathit{valid}$ bit used
for ensuring recoverability. A request that has not experienced a crash, has its $\mathit{valid}$
bit equal to 1, whereas at recovery time, this bit is reinitialized to the value 0. 
At recovery time, this bit is used to disallow a combiner to re-execute a request 
that has already been executed before the crash.
A request whose $\mathit{valid}$ bit is equal to $1$ is called {\em valid}.

\PBcomb\ maintains two records of type $\mathit{StateRec}$
in array $\mathit{MemState}$. It uses them to store copies of the object's state.  
The current state of the implemented object is stored in the element of $\mathit{MemState}$
indexed by the variable $\mathit{MIndex}$.
Each record of type $\mathit{StateRec}$ comprises a field $\mathit{st}$ storing the object's state,
and two arrays.
The first, $\mathit{ReturnVal}$, stores, for each thread, a response 
for the last request initiated by the thread.
The second is the $\mathit{deactivate}$ bit vector.

To reduce the synchronization cost, the implementation of the lock in \PBcomb\ 
is different than in existing combining protocols~\cite{FK12ppopp,FK17opodis,HIST10,KSW18}. 
\PBcomb\ uses an integer shared variable $\mathit{Lock}$:
an odd value stored in it indicates that the lock is taken, 
whereas an even value indicates that the lock is free. 
Implementing the lock in this way, 
allows a thread $\mathit{q}$ to wait on line~\ref{alg:pbcomb:busy_wait}, each time it executes it,
only for the thread $\mathit{p}$ that was the current 
combiner the last time $\mathit{q}$ accessed $\mathit{Lock}$. 
Moreover, $\mathit{q}$ can leave the entry-section without executing \CAS, 
if it discovers that its request has been served. Additionally, 
for each set of combined operations, a single successful \CAS\ is executed. 
Lock implementations in which
every thread should wait its turn to enter the critical-section 
before it leaves the entry-section (e.g., that in~\cite{SFW15}) 
may negatively impact performance.

A thread $\mathit{p}$ starts by recording its request, 
and (the reversed value of) its activate bit, in $\mathit{Request[p]}$. 
Next, it checks if the lock is acquired and if not, it tries to acquire it by executing 
a \CAS. If it succeeds, 
$\mathit{p}$ becomes the combiner and starts executing the {\em combiner code} 
(lines~\ref{alg:pbcomb:combiner:flip_index}-\ref{alg:pbcomb:combiner:unlock}). 
Every thread $\mathit{q}$ that does not become the combiner 
busy waits until the current combiner has released the lock.  
Then, $\mathit{q}$ checks whether its request has been served;
this is true when $\mathit{q}$'s activate and deactivate bits are equal. 
If so, $\mathit{q}$ returns its response value.
The remaining threads contend again for the lock.

In the combining code, 
a combiner $\mathit{p}$ chooses among the two $\mathit{StateRec}$ records of $\mathit{MemState}$
the one, $\mathit{r}$, that is not indexed by $\mathit{MIndex}$, to use for serving requests.
Then, in line~\ref{alg:pbcomb:combiner:state_copy}, it copies the current state of the object 
into $\mathit{r}$. 
Next, it executes a for loop ({\em simulation phase}), 
where for each thread $\mathit{q}$: 
If there is an active, valid request by  $\mathit{q}$,
$\mathit{p}$ a) applies the request
using $\mathit{r}$, ii) records the response into 
the appropriate element of the $\mathit{ReturnVal}$ array stored in $\mathit{r}$, 
and iii) changes $\mathit{Deactivate[q]}$ in $\mathit{r}$ 
to make it equal to its activate bit. 
As soon as $\mathit{p}$ completes the simulation phase, 
it changes $\mathit{MIndex}$ to index $\mathit{r}$ 
and unlocks $\mathit{Lock}$ 
giving up its combining role. 

We say that $\mathit{req}$ has {\em taken effect} at some point $\mathit{t}$,
if a combiner $\mathit{p}$ a) has read $\mathit{req}$ in $\mathit{Request}$ by $\mathit{t}$, b) has performed line~\ref{alg:pbcomb:apply}
for $\mathit{q}$ {\em applying} $\mathit{req}$,
and  c) has executed line~\ref{alg:pbcomb:sync_MIndex} by $\mathit{t}$.
We discuss the correctness of \PBcomb\ in~\cite{FKK21}. 
To comply with persistence principle~\ref{low number},
\PBcomb\ stores a number of its variables in DRAM. This results in improved performance. 
The algorithm can be easily modified to work correctly even if
all data are stored in non-volatile memory.

\noindent
{\bf Persistence.}
When \Recover($\mathit{func}$, $\mathit{args}$, $\mathit{seq}$) is called for a thread $\mathit{p}$, 
$\mathit{p}$ first executes line~\ref{alg:pbcomb:recovery:update-Request}, where it recovers its own entry 
in $\mathit{Request}$. This is necessary to appropriately set $\mathit{p}$'s $\mathit{activate}$ and $\mathit{valid}$ bits. 
This way a combiner is disallowed to re-execute (or to avoid execute) $\mathit{p}$'s request by seeing
an erroneous initial value in $\mathit{p}$'s $\mathit{activate}$ bit after a crash. 
Next,
$\mathit{p}$ checks whether 
the last bit of $\mathit{seq}$ is the same as $\mathit{MemState[MemIndex].deactivate[p]}$. If yes,
then the request has been executed and its response is returned. 
Otherwise, 
$\mathit{p}$ re-invokes \PBcomb($\mathit{func}$, $\mathit{args}$, $\mathit{seq}$). 
Recall that we assume that the system calls \Recover, for each recovered operation, with the same parameters as \PBcomb.

We next explain the role of each of the persistence instructions of \PBcomb.
If the \pwbi\ instructions of lines~\ref{alg:pbcomb:persist_MemState} and~\ref{alg:pbcomb:persist_MIndex}
do not exist, a thread will have no way, at recovery time, to find the current state of the object,
or its response.
Additionally, a \pfencei\ (line~\ref{alg:pbcomb:sync_MemState}) must exist between
these \pwbi s:
Assume that a crash occurs just after
$\mathit{MIndex}$ has been persisted (line~\ref{alg:pbcomb:persist_MIndex}).
If no \pfencei\ exists between the \pwbi s of lines~\ref{alg:pbcomb:persist_MemState} and~\ref{alg:pbcomb:persist_MIndex}, 
then the \pwbi\ on $\mathit{MemState[MIndex]}$ could 
be delayed and thus, the contents of $\mathit{MemState[MIndex].st}$ may be partially persisted, 
at the time of the crash. (Note that $\mathit{MemState[MIndex]}$ may be stored in more than one cache line.)
Consider a request $\mathit{req}$ by a thread $\mathit{q}$ that has been served using $\mathit{MemState[MIndex].st}$ 
and assume that the part of the state reflecting $\mathit{req}$'s updates, has not been
persisted before the crash. 
Assume that the $\mathit{Deactivate[q]}$ bit of $\mathit{MemState[MIndex]}$ has been persisted 
before the crash.
Then, at recovery time, the value of the last bit of $\mathit{seq}$ is the same as 
$\mathit{MemState[MIndex].Deactivate[p]}$, and 
$\mathit{req}$ responds (line~\ref{alg:pbcomb:recovery:rec-applied}),
thus violating durable linearizability.

Assume now that the \psynci\ of line~\ref{alg:pbcomb:sync_MIndex} is missing. Consider an
execution where 
a request $\mathit{req}$ by a thread $\mathit{q}$ has been applied by a combiner $\mathit{p}$.
Assume that after $\mathit{p}$ releases the lock, $\mathit{q}$
 responds for $\mathit{req}$ (line~\ref{alg:pbcomb:return}). 
Then, if a crash occurs, it may happen
that $\mathit{MIndex}$ has not yet been persisted before
this crash. At recovery, the 
object returns in some earlier state, thus violating durable linearizability.

The persistence of $\mathit{ReturnVal}$ and $\mathit{deactivate}$, and the use of 
$\mathit{seq}$, are required only to ensure detectability. 
Consider a request $\mathit{req}$ (initiated by thread $\mathit{q}$) that has taken effect, and assume that 
the system crashes before $\mathit{req}$ completes, i.e., $\mathit{q}$ should be able to find the response of $\mathit{req}$ 
at recovery time. 
If $\mathit{ReturnVal[q]}$ was not persisted, $\mathit{q}$ could not find $\mathit{req}$'s response 
at recovery time, violating detectability. 

The persistence of $\mathit{deactivate}$, as well as the use of $\mathit{seq}$, 
allow $\mathit{q}$ to determine whether $\mathit{req}$ took effect before a crash.
In line~\ref{alg:pbcomb:combiner:if_apply}, \PBcomb\ compares $\mathit{q}$'s activate and deactivate bits 
to determine whether $\mathit{req}$ is still active.
Note that persisting the activate bits (in addition to $\mathit{deactivate}$)
would not be enough to determine whether $\mathit{req}$ was still active when a crash occurred.
Assume that a thread $\mathit{q}$ executes two consecutive requests, $\mathit{req_1}$ and $\mathit{req_2}$ 
of the same type with the same arguments, and the system crashes just 
before $\mathit{req_1}$ returns. 
Thread $\mathit{p}$ cannot distinguish this situation from the case where $\mathit{req_2}$
has just been invoked, as these 
two bits will
be equal in both cases. 
However, in the first case, $\mathit{q}$ should return the response
of $\mathit{req_1}$, whereas in the second, it should re-execute $\mathit{req_2}$. 
To be able to distinguish these two cases, additional (system) support is required~\cite{BHR20}.
Following previous work~\cite{FriedmanQueue18}, \PBcomb\ makes use of the $\mathit{seq}$ parameter.
To reduce persistence cost, \PBcomb\ avoids persisting both $\mathit{seq}$ and $\mathit{Activate[q]}$. 
As $\mathit{seq}$ is provided by the system, there is no need for \PBcomb\ to persist it; we let its
last bit play the role of the activate bit at recovery, so \PBcomb\ does not  persist 
$\mathit{activate}$. 

Note that in a durably linearizable version of \PBcomb, 
the only field of $\mathit{StateRec}$ that needs to be persisted
is $\mathit{st}$.
This will reduce the number of cache lines that need to be persisted
(by the \pwbi\ of line~\ref{alg:pbcomb:persist_MemState}). 
Moreover, the durable linearizable version of \PBcomb\
has {\em null recovery}~\cite{IMS16},
i.e., no recovery function is necessary. 
Detectable recoverability for \PBcomb\ is further discussed in~\cite{FKK21}.)

\vspace{-0.1in}

\section{Wait-free Recoverable Combining}
\begin{algorithm}[t]
	\setcounter{AlgoLine}{0}
		\removelatexerror
		\scriptsize
		\begin{flushleft}
			\Type\ RequestRec \{  \label{alg:pwfcomb:AnnounceRec-start}\;
			\hspace*{3mm} Function $func$\;
			\hspace*{3mm} Argument $args$\;
			\hspace*{3mm} Bit $activate$\;
			\hspace*{3mm} Bit $valid$\;		
\nonl		\}							\label{alg:pwfcomb:AnnounceRec-end}\;
			
			\Type\ StateRec \{\;
			\hspace*{3mm} State $st$\;
			\hspace*{3mm} ReturnValue $ReturnVal[0..n-1]$\;
			\hspace*{3mm} Bit $Deactivate[0..n-1]$\;
			\hspace*{3mm} Bit $Index[0..n-1]$ \label{alg:pwfcomb:index-array}\;
			\hspace*{3mm} $\{0,1,..,n-1\}$ $pid$\;
\nonl   	\}
			
			\vspace*{1mm}
			\com Shared non-volatile variables:\;
			StateRec $MemState[0..n][0..1]$, initially $\langle \bot, \langle \bot,\ldots,\bot\rangle, \langle 0,\ldots,0\rangle, \langle 0,\ldots,0\rangle, 0\rangle$\;
			StateRec *$S := \&MemState[n][0]$\;
			\vspace*{1mm}
			\com Shared volatile variables:\;
			RequestRec $Request[0..n-1]$, initially $\langle\langle\bot,\bot,0,0\rangle, \ldots, \langle\bot,\bot,0,0\rangle\rangle$ \label{alg:pwfcomb:announce-array}\;
			Integer $Flush[0..n-1]$, initially $\langle 0, \ldots, 0\rangle$\;
			Integer $CombRound[0..n-1][0..n-1]$, initially $\langle\langle 0, \ldots, 0\rangle, \ldots, \langle 0, \ldots, 0\rangle\rangle$
		\end{flushleft}

		\vspace*{2mm}


		\myproc{{ReturnValue {\footnotesize \PWFcomb}(Function $func$, Argument $args$,\\ Integer $seq$)}}{
	   		\tcp{Announce request}
			$Request[p] := \langle func, args, 1-Request[p].activate,1 \rangle$ \label{alg:pwfcomb:announce}\;
			Backoff() \label{alg:pwfcomb:backoff}\;
			\KwRet \PerformRequest()
		}			

		\vspace*{2mm}

		\myproc{ReturnValue {\footnotesize \Recover}(Function $func$, Argument $args$,\\ Integer $seq$)}{
\nl			$Request[p]:= \langle func, args, seq \mod 2, 1\rangle$\;
			\tcp{if request is not yet applied}
			\uIf {$S \rightarrow Deactivate[p] \neq seq \mod 2$ } {	\label{alg:pwfcomb:recovery:rec-not-announced}
				\KwRet\ \PerformRequest()		\label{alg:pwfcomb:recovery:re-execute-rec}
			}
			\tcp{request is applied}
			\KwRet $S \rightarrow ReturnVal[p].ret$  					\label{alg:pwfcomb:recovery:rec-applied}
		}
	\caption{\PWFcomb\ - Code for thread $p \in \{0,\ldots, n-1\}$}
	\label{alg:pwfcomb-types}
\end{algorithm}

\begin{algorithm}[t]
	\setcounter{AlgoLine}{0}
		\removelatexerror
		\scriptsize
		\myproc{ReturnValue {\footnotesize \PerformRequest}()}{
\nl			StateRec *$lsPtr$ \;
\nl			Integer $lval$ \;
\nl			\For{$l\gets1$ \KwTo $2$} {                                                 	\label{alg:pwfcomb:two_attempts}
\nl				$lsPtr := \LL(S)$ \;                    \label{alg:pwfcomb:ll}
\nl				Bit $ind := lsPtr\rightarrow Index[p]$ \;						\label{alg:pwfcomb:index-read}
\nl				$MemState[p][ind] := $ *$lsPtr$      \tcp*{copy current state} 		\label{alg:pwfcomb:copy_state}
				$MemState[p][ind].pid := p$\;
				$lval := Flush[lsPtr \rightarrow pid]$\;			\label{alg:pwfcomb:Flush:read}
				\lIf {$lval \mod 2 = 0$} {
					$lval := lval +1$
				}
				\lElse {
					$lval := lval+2$								\label{alg:pwfcomb:lval:init}
				}								
				\lIf {$\VL(S) = \False$} {\Continue}                                    \label{alg:pwfcomb:vl}
				\For{$q\gets0$ \KwTo $n-1$} {                                           \label{alg:pwfcomb:apply_all}
					\tcp{if $q$ has a request that is not yet applied}
					\uIf {$Request[q].valid=1$ \And\ $Request[q].activate \neq MemState[p][ind].Deactivate[q]$} { \label{alg:pwfcomb:if_apply}
						{\bf apply} $Request[q].func$ with $Request[q].args$ on $MemState[p][ind].st$ \label{alg:pwfcomb:apply} \;
						{\bf compute} return value, $returnVal$ \;
						$MemState[p][ind].ReturnVal[q] := returnVal$  \label{alg:pwfcomb:response} \;
						$MemState[p][ind].Deactivate[q] := Request[q].activate$ \label{alg:pwfcomb:deactivate} \;
						$CombRound[p][q] := lval$ \label{alg:pwfcomb:CombRound:set}
					}
				}									\label{alg:pwfcomb:apply_all-end}
				\uIf {$\VL(S) = \True$}{														\label{alg:pwfcomb:vl-2}
					$MemState[p][ind].Index[p] := 1 - MemState[p][ind].Index[p]$ \; \label{alg:pwfcomb:index-update}
					\pwb{\&$MemState[p][ind]$} \;					\label{alg:pwfcomb:MemState:pwb}			
					\pfence{} \;							\label{alg:pwfcomb:MemState:psync}
					
					$Flush[p]:= lval$ \;										\label{alg:pwfcomb:Flush:set}
					\tcp{Try to change $S$ content}
					\uIf {$\SC(S, \&MemState[p][ind]) = \True$} {          		\label{alg:pwfcomb:sc}
						\pwb{\&$S$} \;						\label{alg:pwfcomb:S:pwb}
						\psync{} \;						\label{alg:pwfcomb:S:psync}
						$\CAS(\&Flush[p], lval, lval+1)$ \;		\label{alg:pwfcomb:Flush:reset}
						\KwRet $S \rightarrow ReturnVal[p]$				
					}
					BackoffCalculate();
				}
			}
			
\nl			$lsPtr := S$ \;
\nl			$lval := Flush[lsPtr \rightarrow pid]$\;					\label{alg:pwfcomb:Flush:read-2}
\nl			\uIf {$lval \mod 2 = 1$ \And $lval = CombRound[lsptr\rightarrow pid][p]$} {\label{alg:pwfcomb:flush:helper:read}
\nl				\pwb{\&$S$} \;								\label{alg:pwfcomb:S:pwb-2}
\nl				\psync{} \;								\label{alg:pwfcomb:S:psync-2}
\nl				$\CAS(Flush[p], lval, lval+1)$\label{alg:pwfcomb:flush:helper:reset}
			}
\nl			\KwRet $S \rightarrow ReturnVal[p]$						\label{alg:pwfcomb:end}
		}
	\caption{\PWFcomb\ - Code for thread $p \in \{0,\ldots, n-1\}$}
	\label{alg:pwfcomb-funcs}
\end{algorithm}

In \PWFcomb\ (Algorithms~\ref{alg:pwfcomb-types} and~\ref{alg:pwfcomb-funcs}), all threads pretend to be the combiner: they copy the 
state of the object locally and use this local copy to apply
all active requests they see announced. 
Then, each of them attempts to change a pointer, $\mathit{S}$, to point to its own local copy 
using \SC. 
If  a thread $\mathit{p}$ manages to do so, then $\mathit{p}$ is indeed the thread that acted as the combiner.
\PWFcomb\ borrows ideas from the universal constructions
in~\cite{FK11spaa,FK14,H93}, and can serve as a highly efficient, persistent version 
of these algorithms.

Similarly to \PBcomb, \PWFcomb\ uses a $\mathit{Request}$ array 
and a $\mathit{StateRec}$ record that contains 
the state of the implemented object,
the array of $\mathit{deactivate}$ bits and the array $\mathit{ReturnVal}$.
\PWFcomb\ maintains $2$ records of type $\mathit{StateRec}$ 
for each thread (in addition to two dummy such records, needed for correct initialization). 
This is necessary as each thread pretends to be the combiner:
each thread has to use two $\mathit{StateRec}$ records of its own, to copy the state of
the object locally. Because of this, achieving recoverability is 
more complicated in \PWFcomb\ than in \PBcomb. 
The array $\mathit{Index}$ of a $\mathit{StateRec}$ record aims at coping with some of these complications.
To ensure that persistence principles~\ref{low number} and~\ref{low cost} (Definition~\ref{pp})
are respected, we use the $\mathit{flush}$ integer and the $\mathit{CombRound}$ array (more details below).

To execute a request $\mathit{req}$, a thread $\mathit{p}$ announces $\mathit{req}$ 
and calls \PerformRequest\ 
to serve active requests (including its own).  
In \PerformRequest, $\mathit{p}$ 
reads $\mathit{S}$ (line~\ref{alg:pwfcomb:ll}) and decides 
which of the two $\mathit{StateRec}$ records in its pool it will use (line~\ref{alg:pwfcomb:index-read});
this information is recorded in $\mathit{Index[p]}$ of the $\mathit{StateRec}$ record pointed to by 
the value of $\mathit{S}$ that $\mathit{p}$ read. Next, it makes a local copy of the $\mathit{StateRec}$ record pointed to by $\mathit{S}$ (line~\ref{alg:pwfcomb:copy_state}).
Making a local copy of the state is not atomic, thus
$\mathit{p}$ validates on  line~\ref{alg:pwfcomb:vl} that its local copy is consistent.
Then, $\mathit{p}$ proceeds to the simulation phase (which is similar to that of \PBcomb). 
Afterwards,
it executes an \SC\ 
in an effort to update $\mathit{S}$ to  point to the $\mathit{StateRec}$ on which it was working (line~\ref{alg:pwfcomb:sc}).
Since \PWFcomb\ builds upon and extends \PSim~\cite{FK11spaa,FK14} (see Section~\ref{sec:rel}), 
proving its correctness (in the absence of failures)
follows similar arguments as for \PSim~\cite{FK14}.

The recovery function of \PWFcomb\ is the same as that of \PBcomb.
We focus on the persistence challenges that arise due to the recycling of $\mathit{StateRec}$ records. 
The fact that a thread has two $\mathit{StateRec}$ records and uses them alternatively,
ensures that no thread ever performs active requests on the $\mathit{StateRec}$ record pointed to by $\mathit{S}$.
Thus, threads that read the currently active state see consistent data. 
A variable for each thread $\mathit{p}$, points to the $\mathit{StateRec}$ record that $\mathit{p}$ will
use next. We store these variables into the $\mathit{Index}$ array of $\mathit{StateRec}$,
so that $\mathit{p}$ persists  them together with the $\mathit{StateRec}$ it uses
(at lines~\ref{alg:pwfcomb:MemState:pwb}-\ref{alg:pwfcomb:MemState:psync}),
in accordance to persistence principle~\ref{consecutive}.  
Persisting $\mathit{Index}$ is necessary, since otherwise 
the following bad scenario may happen.
Assume that one of the \SC\ instructions executed by $\mathit{p}$
is successful and let $\mathit{ind}$ be the value that $\mathit{p}$ reads on line~\ref{alg:pwfcomb:index-read},
before the execution of $\mathit{SC}$. 
Assume also that the system crashes after $\mathit{p}$ persists the new value of $\mathit{S}$ and completes. 
Upon recovery, $\mathit{p}$ discovers that its last request has been completed and invokes
a new request. Then, it may happen that $\mathit{p}$ chooses again the same record $\mathit{MemState[p][ind]}$,
and start serving new requests on the current state of the object.
Other active threads may, thus, read  (on line~\ref{alg:pwfcomb:copy_state}) inconsistent data.

Before a thread $\mathit{p}$, that has initiated a request $\mathit{req}$, responds, 
$\mathit{p}$ must persist the value of $\mathit{S}$. 
This should be done independently of whether $\mathit{p}$ has successfully executed the \SC\ of line~\ref{alg:pwfcomb:sc},
for the following reason. 
Assume that $\mathit{p}$ responds for $\mathit{req}$ without persisting $\mathit{S}$ and then the system 
crashes. Upon recovery, 
$\mathit{S}$ will point to a $\mathit{StateRec}$ corresponding to some previous state of the object
than that in which $\mathit{p}$ read $\mathit{req}$'s response. This could violate detectable recoverability. 
For the same reason, executing just the \pwbi\ of line~\ref{alg:pwfcomb:S:pwb} 
(or~\ref{alg:pwfcomb:S:pwb-2}) is not enough
and the \psynci\ of line~\ref{alg:pwfcomb:S:psync} (or~\ref{alg:pwfcomb:S:psync-2}) is also needed. 

Experiments showed that having all threads performing a \pwbi\ and a \psynci\ to persist 
the contents of $\mathit{S}$ before completing, results in high persistence cost. 
This is not surprising as this approach violates persistence principles~\ref{low number}
and~\ref{low cost}. 
To respect these principles, arrays $\mathit{Flush}$ and $\mathit{CompRound}$ are used. 
$\mathit{Flush}$ has one entry for each thread and it is used to indicate whether $\mathit{S}$ 
has already been persisted or not, as described below.
Consider a combiner $\mathit{p}$ that successfully updates $\mathit{S}$
(executing the \SC\ of line~\ref{alg:pwfcomb:sc}).
Before executing the corresponding \SC, 
$\mathit{p}$ changes 
$\mathit{Flush[p]}$ to an odd value 
(lines~\ref{alg:pwfcomb:Flush:read}-\ref{alg:pwfcomb:lval:init} and \ref{alg:pwfcomb:Flush:set}).
Then, after (updating and) persisting $\mathit{S}$, 
$\mathit{p}$ updates $\mathit{Flush[p]}$ to an even value 
(line~\ref{alg:pwfcomb:Flush:reset}), indicating that this change of $\mathit{S}$ has already been 
persisted.
All other threads persist $\mathit{S}$ only if $\mathit{Flush[p]}$ contains an odd value 
(first condition of line~\ref{alg:pwfcomb:flush:helper:read}), in which case, they update 
$\mathit{Flush[p]}$ to the next even value (line~\ref{alg:pwfcomb:flush:helper:reset}).
The use of array $\mathit{CombRound}$ allows a thread $\mathit{q}$ to persist only 
the change of $S$ performed by the combiner $\mathit{p}$ that served $\mathit{q}$'s request, as follows.
For each thread whose request it serves, $\mathit{p}$ stores in its row of $\mathit{CombRound}$  the 
odd value that its $\mathit{Flush[p]}$ integer has when it executes the corresponding \SC. 
Thread $\mathit{q}$ persists $\mathit{S}$ only if $\mathit{Flush[p]}$ contains $\mathit{CombRound[p][q]}$ 
(second condition of line~\ref{alg:pwfcomb:flush:helper:read}).
These techniques contribute to preserving persistence principles~\ref{low number} and~\ref{low cost}.
To maintain persistence principle~\ref{low number}, we store both $\mathit{Flush}$ and
$\mathit{CombRound}$ in volatile memory.

\vspace*{-.3cm}
\section{Recoverable Data Stuctures}
\label{sec:ds}

Here, we provide summaries of our recoverable data structures.
More details and pseudocodes are provided in~\cite{FKK21}.

\noindent
{\bf PBStack and PWFstack.} The stack is implemented as a linked list of nodes. 
Since the stack has a single point of synchronization, the state of the
stack maintained by our algorithms is just the value of $\mathit{top}$, 
the pointer pointing to the topmost element of the stack.
A combiner  $\mathit{p}$ copies the appropriate element of $\mathit{MemState}$, 
reads the current value of $\mathit{top}$ from it, and
serves the active requests using the element, $r$, of $\mathit{MemState}$
that $\mathit{p}$ has chosen to work on. 
To serve a \Push, $\mathit{p}$ has to additionally allocate a new node
and set the next pointer of it to point to the value of $\mathit{top}$ it read.
The combiner persists the fields of all newly allocated nodes 
before persisting $\mathit{e}$ (see also memory management below).
The combiner applies elimination~\cite{HSY04} to pair off
concurrent \Push\ and \Pop\ operations without accessing the state
of the object.  This has small positive impact in performance (Figure~\ref{figs:pstack}).

\noindent
{\bf PBQueue.}
\PBqueue\ uses a singly-linked list to store the nodes of the queue.
To increase parallelism and enhance performance, 
we do not employ \PBcomb\ in an automatic way. 
We rather utilize two instances of \PBcomb, 
one to synchronize the enqueuers ($\mathit{I_E}$), and another 
to synchronize the dequeuers  ($\mathit{I_D}$); thus, combiners of $\mathit{I_E}$ 
serve only enqueue requests, 
while combiners of $\mathit{I_D}$ serve only dequeue requests.
This results in increased parallelism: enqueues are executed concurrently
with dequeues (but sequentially to other enqueues). 
$\mathit{I_E}$ stores just the queue's tail pointer in the $st$ field of its $\mathit{StateRec}$ records,
whereas $\mathit{I_D}$ stores just the head pointer.
\Enqueue\ and \Dequeue\ operations add and remove nodes directly to and from the 
linked list that implements the queue. 
The first list node always plays the role of a dummy node.

The persistence scheme of \PBcomb\ guarantees that the head and the tail 
of the queue are persisted. 
\PBqueue\ also persists the 
modifications performed by the combiners on the nodes of the linked list. 
This is necessary, since otherwise, these modifications will not survive 
after a crash, which may result in an inconsistent state and 
violate durable linearizability.
A \Dequeue\ only updates the head of the simulated queue 
and does not modify the nodes of the linked list. 
Therefore, the effects of a dequeue combiner on the simulated state of the queue are 
correctly persisted by the dequeue instance of \PBcomb.
However, there is a subtlety that needs to be addressed 
regarding the nodes of the linked list that can be removed by the dequeue combiners.
An enqueue combiner simulates the active \Enqueue\ requests by directly modifying 
the nodes of the linked list and then persisting these modifications.
Thus, if no care is taken, a dequeue combiner may remove list nodes that have been 
appended by an active enqueue combiner but not yet persisted.
This may jeopardize detectable recoverability.
To address this, \PBqueue\ disallows dequeue combiners 
to remove any node from the linked list that has not yet been persisted. 
It achieves this by using a shared volatile variable $\mathit{oldTail}$.
An enqueue combiner updates $\mathit{oldTail}$ to point to the last node of the queue
after it persists its changes and before releasing the lock. 
A dequeue combiner removes nodes from the linked list up until $\mathit{oldTail}$.

\noindent
{\bf PWFQueue.}
\PWFqueue\ combines ideas from \PBqueue\ and 
\SimQueue~\cite{FK11spaa,FK14}. 
As in \PBqueue, the queue is implemented as a singly-linked list and
\PWFqueue\ uses two instances of \PWFcomb\ ($\mathit{I_E}$ and $\mathit{I_D}$) to synchronize
the enqueuers and the dequeuers. 
A thread executing an \Enqueue\ will also try to serve \Enqueue s by other 
enqueuers. It does so by creating a local list of new nodes that will eventually be 
appended to the current state of the queue. 
So, at some point in time, the linked list implementing the queue may be comprised of two parts.
To ensure consistency, all threads 
perform the linking of these parts before they proceed to serve requests.
Also, the state maintained by $\mathit{I_E}$ is now comprised of three 
pointers to support the linking of the two parts of the list. 

Regarding persistence, 
some subtleties arise from the necessity to connect the two parts of the linked list representing
the queue. 
Before updating the queue's tail, an enqueuer $\mathit{p}$ has to persist 
the pointers needed to connect the two parts of the linked list,
i.e., the current tail of the queue and the pointer to the first node
of its local list. 
If it does not do so, then the system may crash just after 
$\mathit{p}$ updates the tail (and before it connects the two parts of the linked list), 
in which case its local copy is lost and durable linearizability may be jeopardized.
Additionally, an enqueuer that connects the linked list, has to persist the new values 
of the node it updated (i.e., its pointer to the next element of the linked list).
Although dequeuers also help connecting the two parts of the list,
it is enough to persist only the head of the queue.

The code and a more detailed description of \PBqueue\ and \PWFqueue\ are 
provided in~\cite{FKK21}.

\noindent
{\bf PBHeap.}
\PBheap\ is a persistent bounded min-heap implementation based on \PBcomb.
The state stored in $\mathit{StateRec}$ is the array of heap elements and two integers 
identifying the bounds of the heap.
\PBheap\ supports the operations \HGet, \HInsert, and \HDelete. 
It employs a single instance of \PBcomb\ and 
is implemented by enhancing a sequential heap implementation 
with the code of \PBcomb.

\noindent 
{\bf Memory Management.}
For ensuring persistence in allocating new stack or queue nodes,
we follow a standard technique~\cite{RC+19,CFR18,DFC} where each thread $\mathit{p}$ 
pre-allocates a fixed-size memory chunk in NVMM,
and reserves nodes from this chunk. 
Whenever this chunk is exhausted a new memory chunk is allocated by $\mathit{p}$.
If a garbage collection mechanism (for collecting nodes) is not used, 
whenever $\mathit{p}$ serves as the combiner, it gets nodes 
in consecutive memory addresses (to comply with the persistence principles). 

For garbage collection, in \PBqueue, each thread $\mathit{p}$ has its own free list and places
there nodes it removes when acting as combiner. It does so, 
after 
causing the removal of these nodes to take effect.
Whenever $\mathit{p}$ needs to reserve nodes while its free list is not empty,
it uses nodes from this list. Note that this does not ensure 
persistence principle~\ref{consecutive}, as the nodes in its free list
may belong to chunks of other threads. 
We were able 
to implement an efficient garbage collection scheme for \PBstack\ (by exploiting its semantics). 
We maintain a single free list for all threads, implemented as a stack ({\em recycling stack}). 
Whenever $\mathit{p}$ needs to reserve a node, 
it pops a node from the recycling stack.
This way recycled nodes are re-inserted in the implemented data structure 
in the same order as they have originally been reserved from the memory chunk. 
This complies with persistence principle~\ref{consecutive}.  

To support garbage collection in \PWFstack, 
we extend the scheme described for \PBqueue\ with the simple validation scheme of~\cite{BY20},
which disallows a thread to access nodes that have already been placed in a free list.
For \PWFqueue, a solution would be more complicated, due to the fact that there
may be two parts that comprise the state of the queue.
We have left this for future work.

\vspace*{-.3cm}
\section{Performance Evaluation}
\label{sec:perf}
We evaluate our algorithms on a 48-core machine (96 logical cores) consisting of 
2 Intel Xeon Platinum 8260M processors with 24 cores each. 
Each core executes two threads concurrently. Our machine is equipped with a 1TB Intel Optane 
DC persistent memory (DCPMM) and the system is configured in AppDirect mode.
We use the 1.9.2 version of the \textit{Persistent Memory Development Kit}~\cite{PMDK-web}, which
provides the \pwbi\ and \psynci\ persistency instructions.
An $x86\_64$ store fence instruction is used for implementing a \pfencei\ operation.
The operating system is Linux
(kernel v3.4) and we use \textit{gcc} v9.1.0. 
Threads were bound in all experiments following a scheduling policy
which distributes the running threads evenly across the machine's NUMA nodes~\cite{FK12ppopp,FK14}.
For our experiments, we simulate an \LL\ on an object $O$ with a
read, and an \SC\ with a \CAS\ on a timestamped version of $O$ to avoid the ABA problem.
We executed each experiment 10 times (runs) and display averages.
Each run simulates $10^7$ atomic operations in total, with each of the $n$
threads simulating $10^7/n$ operations. 
In the experiments for the stacks  (queues),
each thread performs pairs of \Push\ and \Pop\ (\Enqueue\ and \Dequeue)
starting from an empty data structure. This experiment is kind of standard~\cite{FK11spaa,FK12ppopp,FK14,FriedmanQueue18,DFC,SP21},
as it avoids performing unsuccessful (and thus cheap) operations. 
We performed also experiments where each thread executed random operations
(50\% of each type), as well as experiments where the data structure was initially populated; 
as they did not illustrate significant differences in the performance trends of
the tested algorithms,
we do not report these experiments.

\begin{figure}[!t]
	\centering
	\includegraphics[width=1\linewidth]{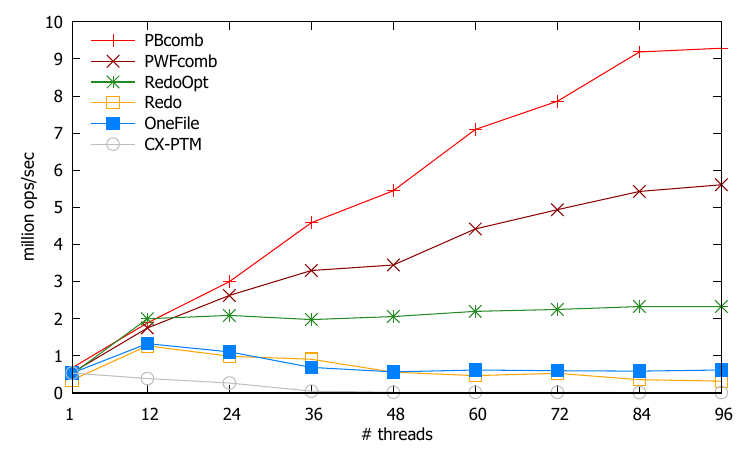}
	\caption{Simulation of a persistent \AtomicFloat\ object on Intel Xeon: throughput.\label{figs:pfam}}	
\end{figure}

\begin{figure}[!t]
	\centering
	\includegraphics[width=1\linewidth]{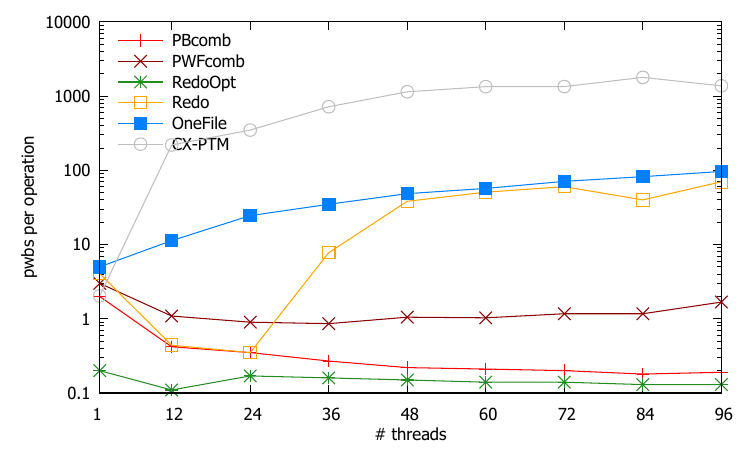}
	\caption{Simulation of a persistent \AtomicFloat\ object on Intel Xeon: \pwbi\ instructions per operation.\label{figs:pfam_pwbs}}
\end{figure}

\begin{figure}[!t]
	\centering
	\includegraphics[width=1\linewidth]{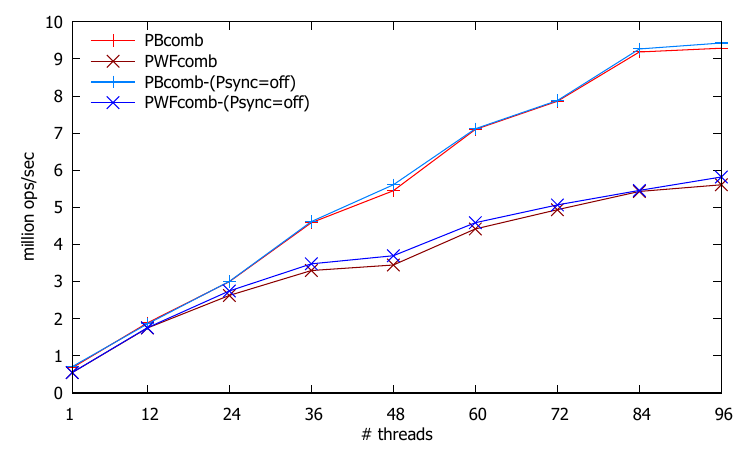}
	\caption{Simulation of a persistent \AtomicFloat\ object on Intel Xeon: throughput with no \psynci\ instructions.\label{figs:pfam_psync_off}}	
\end{figure}

\noindent
{\bf Synthetic Benchmark.}
We first consider a synthetic benchmark (\AtomicFloat) in which
every thread, repeatedly, executes \AtomicFloat($O$, $k$) that reads the value $v$ of $O$ and 
updates it to $v * k$; the thread returns the value read. 
To avoid long runs and unrealistically low number of cache 
misses~\cite{FK12ppopp,FK11spaa,FK17opodis,MCS91}, 
we added a local workload between consecutive executions of atomic operations, 
implemented as a short loop of a random number (maximum 512) of dummy iterations~\cite{FK12ppopp,FK11spaa}. 
In Figure~\ref{figs:pfam}, we compare the performance of \AtomicFloat\ implementations based 
on \PBcomb\ and \PWFcomb\ against state-of-the-art wait-free persistent synchronization techniques:
\OneFile~\cite{RC+19}, \CXPUC~\cite{CFP20eurosys}, \CXPTM~\cite{CFP20eurosys}, and \RedoOpt~\cite{CFP20eurosys},
using the latest version of code for these algorithms 
provided in \cite{redodb_code}.
These algorithms satisfy 
durable linearizability (not detectable recoverability).
Figure~\ref{figs:pfam} shows that \PBcomb\ is more than $4$x faster than \RedoOpt,
which is the fastest among the competitors. 
Also, \PWFcomb\ is more than $2.8$x faster than \RedoOpt.
Figure~\ref{figs:pfam_pwbs} shows that both \PBcomb\  and \PWFcomb\
perform (on average) a small number of \pwbi\ instructions
per operation. Figure~\ref{figs:pfam_psync_off} shows that the impact 
of \psynci\ is negligible in our experiments.
This experiment illustrates that the main persistence cost in our algorithms comes from the \pwbi\ instructions,
and reveals the importance of keeping
the number of \pwbi s (and their cost) low, thus respecting persistence principles~\ref{low number} and~\ref{low cost}, 
when designing persistent synchronization protocols and concurrent data structures. 

\begin{figure}[!t]
	\includegraphics[width=1\linewidth]{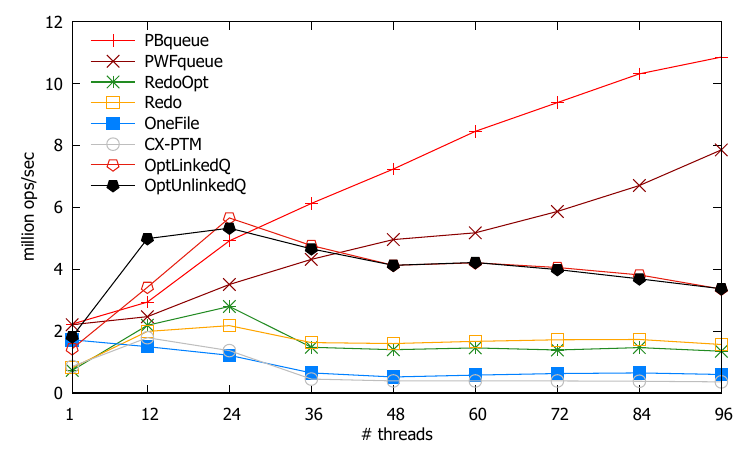}
	\caption{Persistent queue implementations on Intel Xeon: throughput with no \pwbi\ instructions.\label{figs:pqueue_pwbs_off}}
\end{figure}

Note that \PBcomb\ causes almost the same number of \pwbi s as \RedoOpt~\cite{CFP20eurosys}.
\RedoOpt\ uses ideas from \PSim~\cite{FK11spaa}, and thus it employs 
some form of combining. 
Because of this, \RedoOpt\ executes a low number of low cost \pwbi\ instructions.
However, \RedoOpt\ employs a shared queue, 
stored in volatile memory, to impose an order to the executed operations,
which results in high synchronization overhead.
\PBcomb\ achieves the same number of \pfencei s and \psynci s as \RedoOpt\
and does not cause any noteworthy increase to the number of \pwbi s. 
Interestingly, this is achieved at a much lower synchronization cost
(see Figure~\ref{figs:pqueue_pwbs_off}, discussed below). 

\begin{figure}[!t]	
	\includegraphics[width=\linewidth]{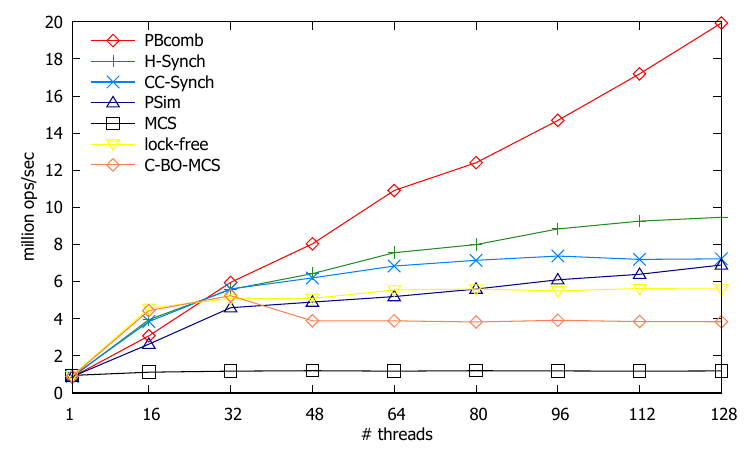}
	\vspace{-0.24in}
	\caption{Throughput of implementations while simulating a volatile \AtomicFloat\ object on AMD Epyc.}
	\label{figs:fam}
\end{figure}

\PBcomb\ performs better than \PWFcomb\ in all experiments.
The main reasons are that 1) the synchronization cost of \PBcomb\ is 
lower than \PWFcomb\ (see Figure~\ref{figs:fam}), and 
2) \PWFcomb\ has higher persistence cost, 
as all threads should ensure that $S$ is persisted before returning.  
These costs are paid to ensure the wait-free property of \PWFcomb.


\begin{figure}[!t]
	\includegraphics[width=1\linewidth]{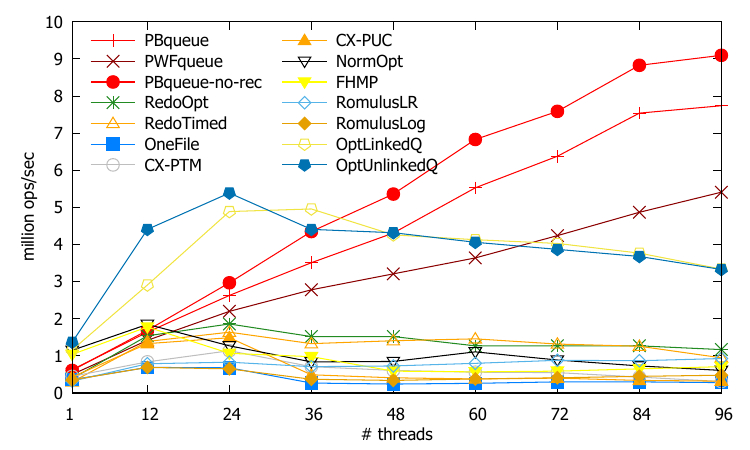}
	\caption{Persistent queue implementations on Intel Xeon: throughput.\label{figs:pqueue}}
\end{figure}
\begin{figure}[!t]
	\includegraphics[width=1\linewidth]{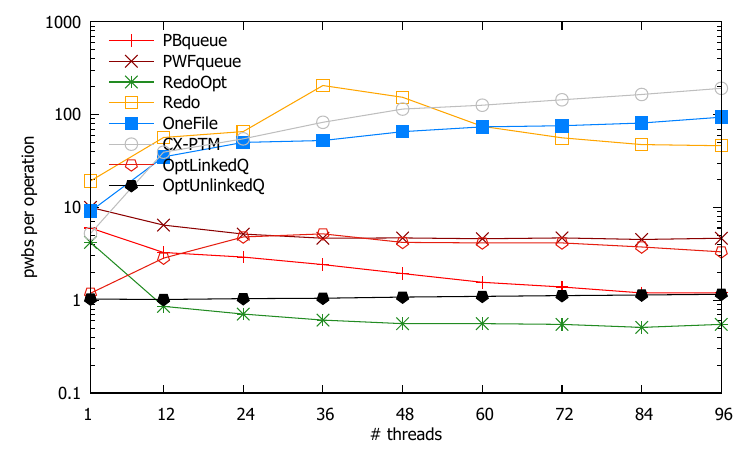}
	\caption{Persistent queue implementations on Intel Xeon: \pwbi\ instructions per operation.\label{figs:pqueue_pwbs}}
\end{figure}

\noindent
{\bf Persistent queues.} 
Figure~\ref{figs:pqueue} compares the performance of \PBqueue\ and \PWFqueue\ with persistent queue implementations
based on the persistence techniques studied in Figure~\ref{figs:pfam}. 
It also compares \PBqueue\ and \PWFqueue\ with the specialized persistent queue implementation 
in~\cite{FriedmanQueue18} (\FHMP), and those recently published in~\cite{SP21} (\LDQ\ and \UDQ), 
as well as the persistent queue implementations  based 
on \CAPSULES~\cite{NormOptQueue19} (\NormOpt), 
and persistent queue implementations based on \Romulus~\cite{CFR18} 
(i.e., RomulusLR and RomulusLog).
Figure~\ref{figs:pqueue} shows that \PBqueue\ achieve 
superior performance by being $2$x faster than the \UDQ, which is the best competitor.
Figure~\ref{figs:pqueue_pwbs} shows the number of \pwbi s
in different queue implementations; trends are similar to Figure~\ref{figs:pfam_pwbs}. 
In Figure~\ref{figs:pqueue_pwbs_off}, 
we have replaced the \pwbi\ instructions with simple NOP operations and we measure the 
throughput of the different algorithms.
The figure shows that the synchronization cost of
\PWFcomb\ and \PBcomb\ is much lower compared to its competitors. 
A comparison of Figure~\ref{figs:pqueue_pwbs} with Figure~\ref{figs:pqueue_pwbs_off} 
shows the performance impact of persistence.

\begin{figure}[!t]
	\begin{subfigure}{\linewidth}
		\includegraphics[width=\linewidth]{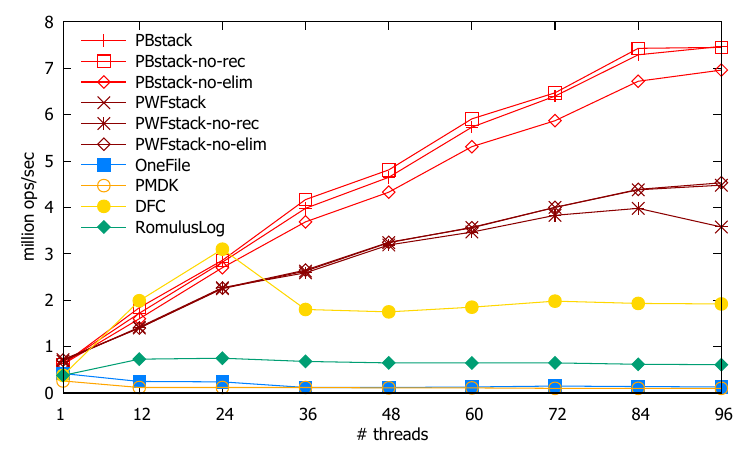}
		\vspace{-0.24in}
		\caption{\centering}
		\label{figs:pstack}
	\end{subfigure}		
	\begin{subfigure}{\linewidth}
		\includegraphics[width=\linewidth]{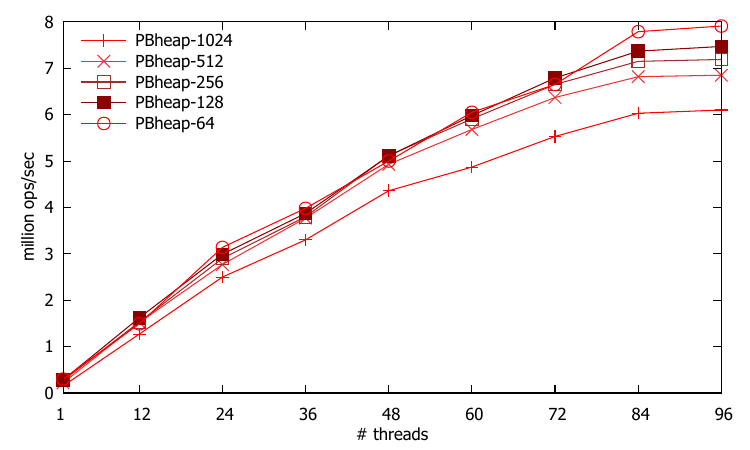}
		\vspace{-0.24in}
		\caption{\centering}
		\label{figs:pheap}
	\end{subfigure}
	\vspace{-0.12in}
	\caption{Experiments on Intel Xeon:
		     (a) throughput of persistent stack implementations and
		     (b) throughput of persistent heap implementations based on \PBcomb\ 
		         for different heap-sizes (64 - 1024).
	}
\end{figure}

\noindent
{\bf Persistent Stacks.}
Figure~\ref{figs:pstack} illustrates that the performance of \PBstack\ and \PWFstack\
is much better than the following algorithms: the persistent stack implementations 
based on \OneFile~\cite{RC+19} and \Romulus~\cite{CFR18}, and a persistent stack 
based on flat-combining (\DFC)~\cite{DFC}, which is the best competitor. 

Similarly to our stack implementations, 
\DFC\ uses an announce array where threads can announce their requests.
In contrast to our algorithms, 
\DFC\ does not avoid the cost of persisting this array. 
\DFC\ has each thread persisting its own element in the announce array.
To ensure durable linearizability, a combiner serves only those requests whose announcements have been persisted. 
This requires an additional mechanism in order for a thread
to inform the combiner that it has persisted its announcement. 
Another  major difference of \DFC\ from our approach
is that in \DFC\ the combiners perform updates directly on the state of the object. 
This introduces several difficulties for achieving persistence when designing the stack. 
Finally, \DFC\ stores the return value for each thread in the announce array. This requires
that the combiner persists each return value separately. 
These design decisions result in high persistence cost and
synchronization overhead, as reflected in the figure. 

\DFC\ applies elimination for reducing its persistence cost. 
However, the \DFC\ design decision of performing updates directly on the shared state
complicates its elimination scheme and its recovery code. 
We also applied elimination to our algorithms. Figure~\ref{figs:pstack}
(comparing the diagrams \PBstack\ and \PWFstack\ with \PBstackNE\ and \PWFstackNE, respectively) 
shows the positive impact of elimination in our stack implementations. As our implementations apply updates
on copies of the state, the positive impact stems mainly 
from reducing their persistence cost (e.g., the number of newly allocated nodes
that need to be persisted).

\noindent
{\bf Memory Management.} 
Diagrams \PBstackNR\ and \PWFstackNR, in Figure~\ref{figs:pstack}, illustrate the impact of
removing the scheme for recycling list nodes in our stacks.  Comparing them with \PBstack\ and \PWFstack\ shows that
our memory management scheme for stacks is very efficient. On the contrary, Figure~\ref{figs:pqueue} 
shows that the performance of \PBqueue\ is negatively affected by the simple recycling scheme for nodes 
we apply in this case (Section~\ref{sec:ds}). 

\noindent
{\bf Persistent Heaps.}
Figure~\ref{figs:pheap} shows the throughput of \PBheap\ for small and medium heap sizes (i.e., $64-1024$ keys).
Initially, the heap is half-full.
To make the experiment realistic, we avoid to have a full (or empty) heap 
by performing an equal number of \HInsert\ and 
\HDelete\ operations. Figure~\ref{figs:pheap} shows that even for heaps of medium size,
the performance of \PBheap\ is good, illustrating 
that more complex persistent data-structures than stacks and queues 
can easily be implemented on top of our algorithms, and perform well when their size is not very large.

\noindent
{\bf Performance in systems with volatile memory.} 
We study the performance of the volatile version of \PBcomb\
in a system without NVMM:
a 64-core AMD Epyc consisting of $2$ Epyc 7501 
processors, which provides $64$ cores ($8$ NUMA nodes, $128$ logical cores);
we saw similar performance behavior on the Intel Xeon machine.
In Figure~\ref{figs:fam}, the synthetic benchmark runs 
using  \HSynch~\cite{FK12ppopp,SynchFramework}, \CCSynch~\cite{FK12ppopp,SynchFramework},
\PSim~\cite{FK11spaa,SynchFramework}, MCS queue spin-locks~\cite{MCS91}, a simple lock-free implementation~\cite{FK11spaa,FK14}, 
and an hierarchical lock (C-BO-MCS)~\cite{cohort_locks}. 
Figure~\ref{figs:fam} shows that a volatile version of \PBcomb\ exhibits much better performance than all other algorithms.
In Table~\ref{table:perf_counters}, we present results for 1)~cache-misses per operation, 
2)~stores on cache-shared locations per operation, and 3)~reads on cache-shared locations per operation. 
(More experiments are provided in~\cite{FKK21}.)

\begin{table}[!t]
	\footnotesize
	\center
	\begin{tabular}{|c|c|c|c|c|}
		\hline
		(per operation) & \Bcomb & \HSynch & \CCSynch & \PSim \\\hline\hline
		cache-misses                           & 2.8     & 6.3     & 5.5      & 10 \\\hline
		stores on cache-lines in shared state  & 0.0012  & 1.0     & 1.0      & 1.0 \\\hline
		reads on cache-lines in shared state   & 0.034   & 1.8     & 1.6      & 1.3 \\\hline
	\end{tabular}
	\caption{Performance counters using \textit{Perf} for $128$ threads.}
	\label{table:perf_counters}
\end{table}

\section{Related Work}
\label{sec:rel}

A lot of work has been devoted to design persistent
transactional memory systems (e.g.,~\cite{VT+11,CC+11-I,CD+14-I,PMDK-web,IKK16,RC+19,BC+20,XIS21,LI+18}).
Such systems often rely on some kind of logging technique employing 
either redo logs~\cite{VT+11,IKK16,RC+19} or undo logs~\cite{CD+14-I,PMDK-web,CC+11-I}. 
Logging causes serious performance penalties as the log
is usually stored in persistent memory. 
Our algorithms avoid logging
to reduce both synchronization and persistence cost.

PMDK~\cite{PMDK-web} 
attempts to reduce the logging cost by aggregating
all updates performed on an object in a single transaction. 
Romulus~\cite{CFR18} follows a different approach for achieving
the same goal.
Romulus comes in two flavors, RomulusLog which is blocking, 
and RomulusLR which supports wait-free read-only transactions
(and blocking update transactions).

OneFile~\cite{RC+19} is a redo-log based persistent transactional system 
whose main characteristic is that its transactions do not maintain read-sets. 
However, it serializes all update transactions and all transactions
(read-only and update) have to help update transactions to complete. 
OneFile  comes in two versions, one lock-free  and another wait-free. 
The wait-free version shares some ideas with \PSim, thus integrating some form of 
combining, but it inherits the helping and logging mechanisms from the 
lock-free version. 
ONLL~\cite{CGZ18} is a log-based persistent universal construction which ensures durable linearizability~\cite{IMS16} and 
lock-freedom.
ONLL performs
one persistent fence for each update operation  and avoids performing
persistence fences for read operations.  

A persistent wait-free universal construction (\CXPUC) and 
a persistent transactional memory system (\CXPTM)
are presented in 
~\cite{CFP20eurosys}. 
Both algorithms are based on the universal construction provided in~\cite{CRP20}.
The algorithms store $2n$ replicas of the data structure in NVMM, 
and use a shared queue, stored in volatile memory,
to impose an order to the executed operations. Threads synchronize using consensus objects
in order to decide the order in which the operations will be applied on the data structure.
A thread chooses one of the persistent copies of the
data structure to work on and may require to execute all operations that precede its operation
in the queue, in order to ensure consistency.
 
\RedoOpt, presented also in~\cite{CFP20eurosys}, is a persistent, durably linearizable, wait-free universal construction
that uses ideas from \PSim\
to achieve lower persistence cost and better performance than \CXPUC\ and \CXPTM.
\RedoOpt\ employs the shared queue used by \CXPUC\ and \CXPTM, and therefore it does not
avoid the synchronization overheads of them.

All these algorithms satisfy weaker consistency than detectable recoverability
ensured by \PBcomb\ and \PWFcomb.

Capsules~\cite{NormOptQueue19} can be used to 
transform concurrent algorithms that use only read and \CAS{} primitives
to their persistent versions.
The programmer has to partition the code into parts, called {\em capsules},
each containing a single \CAS{}.
This \CAS\ has to be replaced
with its recoverable version~\cite{AttiyaBH-PODC2018}.
We use an optimized version of Capsules, which can be applied only to 
\emph{normalized} implementations~\cite{TimnatP-PPoPP2014},
to our experiments, as it achieved better performance.
Recent generic approaches for designing lock-free data structures appear in~\cite{FB+20,FPR21}; 
they are not detectable recoverable and they do not experiment with stacks and queues.

The first hand-tuned durable queues were provided 
in~\cite{FriedmanQueue18}. 
One of them, namely the {\em log-queue}, ensured detectable recoverability,
whereas the other two guaranteed {\em durable linearizability}~\cite{IMS16}
and {\em buffered durable linearizability}~\cite{IMS16}, respectively.
Their design is based on the lock-free queue (\MSQueue)
presented by Michael and Scott~\cite{MSQUEUE96}. 
A recent paper~\cite{SP21} presents hand-tuned durably linearizable queue implementations
that outperform those in~\cite{FriedmanQueue18} and other previous persistent queue implementations. These implementations
are designed based on the observation that minimizing accesses
to flushed content could be beneficial for performance. 
Our experiments show that \PBqueue\ 
outperforms the queues in~\cite{SP21} as the number of threads increases.

\PBcomb\ and \PWFcomb\ borrow and extend ideas from \PSim~\cite{FK11spaa,FK14}, 
a state-of-the-art wait-free practical software combining protocol, which is built upon
the simple idea presented by Herlihy in~\cite{H93}. 
A thread $\mathit{p}$ 
first announces its request and informs other threads that it
has an active request by applying a Fetch\&dAdd instruction on an integer variable
that implements a bit vector. 
Next, it finds out which other requests are active by reading this integer variable,
and applies these requests to a local copy of the simulated
object. Finally, it tries to change the shared pointer
to the simulated object's state to point to this local copy.
Similarly, \PWFqueue\ is the persistent version of \SimQueue. \SimQueue\ allows the enqueuers and
dequeuers to run independently by employing two instances of \PSim.
It also employs a linked list that is comprised of two parts for implementing the queue
and have all threads performing appropriate actions to link these parts before serving requests.

All detectable algorithms we are aware of 
assume some system support to ensure detectability. 
Those in~\cite{BHR20,DFC,AB+new,AB+20} assume that 
for every thread $p$, the system calls the recovery function of the request 
$req$ that $p$ was executing at crash time, with the same 
arguments as $req$. We follow the same assumption in this paper. 
They also assume that $p$ has a non-volatile private variable 
that recoverable operations and recovery functions use for managing
check-points in their execution flow; the system sets the value of
this variable to $\mathit{0}$ just before $p$ initiates the execution of a new request.
Instead, we assume that $p$ has a toggle bit which the system toggles 
each time $p$ invokes a request and passes it as a parameter to the request
(recall that we implement this mechanism through the use of $\mathit{seq}$).
Our algorithms can be adjusted to work using check-pointing variables, 
as in~\cite{BHR20,DFC,AB+new,AB+20}. This may require
to persist private non-volatile variables for each thread,
which is expected to be of low cost~\cite{AB+new}. 
The detectable algorithms in~\cite{FriedmanQueue18,NormOptQueue19} 
assume, as here, the use of a sequence number
which is passed to recoverable operations via their arguments.
Other detectable algorithms~\cite{AttiyaBH-PODC2018} also assume that the system 
persists some of the threads' state. 
Ben-Baruch {\em et al.}~\cite{BHR20} prove that detectability 
cannot be achieved without system support.
Specifically, they prove that for a specific class of objects 
(which include FIFO queues, considered in this paper), 
 any {\em obstruction-free} detectable implementation must receive auxiliary state.

For our experiments, we tested code which is publicly available~\cite{SynchFramework,redodb_code,durablequeues_code,DFC_code},
and we focus on persistent synchronization techniques, transactional memory systems and universal constructions, 
whose experimental platforms provide persistent stack and queue implementations.

\section{Discussion}
\label{sec:concl}

We present \PBcomb\ and \PWFcomb,
highly-efficient recoverable software combining protocols
that are many times faster than state-of-the-art recoverable universal 
constructions and software transactional systems.
We identify three persistence principles, crucial for performance, 
and we illustrate how to make the appropriate design decisions to respect 
them when designing recoverable software combining protocols.
Both \PBcomb\ and \PWFcomb\ can be used to derive recoverable implementations
of any data structure from its sequential implementation. Thus, it is possible 
to develop a software-combining API that automatically transforms 
any data structure to fit our schemes by using a single instance of 
the corresponding algorithm.
Our recoverable implementations of stacks, and the heap implementation, indeed follow this approach,
using a single instance of \PBcomb\ or \PWFcomb.
To increase parallelism and achieve better performance, \PWFqueue\ (and \PBqueue)
employs a similar approach as \SimQueue~\cite{FK11spaa,FK14} utilizing two instances of 
\PWFcomb\ (\PBcomb, respectively).
Although this choice is not fundamentally necessary and 
made the queue implementations more complicated than using a single instance,
it results in superior performance. 

Coming up with a wait-free recoverable heap using \PWFcomb\
is a relatively easy task. We are currently working on this direction,
as well as on implementing a simple garbage collection scheme for \PWFqueue.
We will include the resulting algorithms in future versions of our library.

Software combining restricts parallelism by executing sequentially all requests. 
Thus, \PBcomb\ and \PWFcomb, although applicable, are not necessarily the best choices
for implementing e.g., recoverable tree-like data structures,
where threads may work on different subtrees without interference. 
Experiments for \PBheap\ illustrate that \PBcomb\ and \PWFcomb\ may perform well in this case,
only if the data structure size is small or medium.
In~\cite{AB+new}, we present a generic approach for obtaining efficient recoverable
such data structures, independently of their size, from their concurrent implementations. 

In~\cite{FKR18}, more than one instance of \PSim\ is used to efficiently implement an 
extendible hashing scheme. Using more instances of \PBcomb\ and \PWFcomb\ for efficiently 
implementing recoverable hashing, or recoverable tree-like data structures 
is an interesting open problem.

The performance of state-of-the-art combining protocols \cite{FK12ppopp,HIST10}
is still far from the ideal~\cite{FK17opodis}; 
the ideal performance is measured in~\cite{FK17opodis} by calculating the
time that it takes to a single thread to execute the total number of synchronization requests
(sidestepping the synchronization protocol) and perform the total amount of local work that follows 
its own synchronization requests. 
\cite{FK17opodis} proposes a technique, called Osci, that enables batching of the synchronization requests 
initiated by threads running on the same (oversubscribed) core. It studies the impact on performance 
of this technique, when it is combined with cheap context switching and shows that it is remarkable. 
Osci has performance which is very close to the ideal. 
Klaftenegger {\em et al.}~\cite{KSW18} proposes a technique, similar to {\em futures}~\cite{BH77}, 
where a thread does not block waiting the combiner to serve its request;
it rather executes subsequent computation and may block when it needs to access 
some of the variables that are updated by the request. 
This technique increases parallelism and enhances performance. 
The paper~\cite{KSW18} also focuses on the case where some of the requests do not require any response
and shows that avoiding recording of responses could have a positive impact on performance.
Examining whether
the techniques presented in~\cite{FK17opodis,KSW18} can be extended and combined 
with our results to get more efficient
recoverable protocols 
is a potential path for future work.

A collection of arguments to support correctness of our protocols are provided 
in~\cite{FKK21}.  Using model checking or verification techniques for further checking 
correctness~\cite{BE+17} would be a valid path to consider.

\vspace*{.5cm}
\noindent
{\bf Acknowledgments.} 
This research is supported by the EU Horizon 2020, 
Marie Sklodowska-Curie project with GA No 101031688,
 and by the Hellenic Foundation for Research and Innovation 
(HFRI) under the ``Second Call for HFRI Research 
Projects to support Faculty members and researchers" 
(project number: 3684).
For Eleftherios Kosmas, the research is co-financed by Greece
and the European Union (European Social Fund- ESF) 
through the Operational Programme ``Human Resources Development,
Education and Lifelong Learning" in the context of the project 
``Reinforcement of Postdoctoral
Researchers - 2nd Cycle" (MIS-5033021), 
implemented by the State Scholarships Foundation (IKY).

\bibliographystyle{abbrv}
\bibliography{references}

\newpage
\appendix
\section{Blocking Recoverable Queue}
\label{appendix:PBqueue}
We present \PBqueue, a blocking queue based on \PBcomb.
\PBqueue\ uses a singly linked list $L$ to store the nodes of the queue.
It employs
two instances of \PBcomb, namely $I_E$ and $I_D$,
to synchronize the enqueuers and the dequeuers, respectively;
this results in increased parallelism as it supports enqueues 
that are executed concurrently with dequeues.
To achieve this,
the first node of $L$ plays the role of a dummy node.

\begin{figure}[t]
	\scriptsize		
	
	\begin{flushleft}

		type RequestRec \{ 		 \\	\label{alg:pbqueue:AnnounceRec-start}
		\hspace*{6mm} $\{\Enqueue, \Dequeue\}$ $func$ \\
		\hspace*{6mm} Integer $arg$ \\
		\hspace*{6mm} Bit $activate$ \\
		\hspace*{6mm} Bit $valid$\\
		\}							\label{alg:pbqueue:AnnounceRec-end}		

		type Node \{ \\
			\hspace*{6mm} Data $data$ \\
			\hspace*{6mm} Node *$next$ \\			
		\}

		type EStateRec \{ \\
		\hspace*{6mm} Node *$Tail$ \\
		\hspace*{6mm} ReturnValue $ReturnVal[0..n-1]$ \\
		\hspace*{6mm} Bit $Deactivate[0..n-1]$ \\
		\}

		type DStateRec \{ \\
		\hspace*{6mm} Node *$Head$  \\
		\hspace*{6mm} ReturnValue $ReturnVal[0..n-1]$ \\
		\hspace*{6mm} Bit $Deactivate[0..n-1]$ \\
		\}

		\vspace*{1mm}
		\com Shared non-volatile variable: \\
		Node $DUMMY$, initially $\langle \bot, \bot \rangle$ \\

		\vspace*{1mm}
		\com Shared volatile variables: \\
		\lred{Node *$oldTail$, initially $\&DUMMY$} \\
		\lred{Set$\langle$Node*$\rangle$ $toPersist$, initially $\emptyset$}\\
		
		\vspace*{1mm}
		\com Shared non-volatiles variables used by the \PBqueueEnq\ instance of \PBcomb: \\
		EStateRec $EState[0..1]$, initially $\langle\&DUMMY, \langle\bot,\ldots,\bot\rangle, \langle0,\ldots,0\rangle\rangle$ \\		
		Bit $EIndex$, initially $0$  \\

		\vspace*{1mm}
		\com Shared volatile variables used by the \PBqueueEnq\ instance of \PBcomb: \\
		RequestRec $ERequest[0..n-1]$, initially $\langle\langle\bot,\bot,0,0\rangle, \ldots, \langle\bot,\bot,0,0\rangle\rangle$ \\
		Integer $ELock$, initially $0$ \\
		Integer $ELockVal$, initially $0$ \\

		\vspace*{1mm}
		\com Shared non-volatile variables used by the \PBqueueDeq\ instance of \PBcomb: \\
		DStateRec $DState[0..1]$, initially $\langle\&DUMMY, \langle\bot,\ldots,\bot\rangle, \langle0,\ldots,0\rangle\rangle$ \\		
		Bit $DIndex$, initially $0$  \\

		\vspace*{1mm}
		\com Shared volatile variable used by the \PBqueueDeq\ instance of \PBcomb: \\
		RequestRec $DRequest[0..n-1]$, initially $\langle\langle\bot,\bot,0,0\rangle, \ldots, \langle\bot,\bot,0,0\rangle\rangle$ \\
		Integer $DLock$, initially $0$ \\
		Integer $DLockVal$, initially $0$ \\

	\end{flushleft}	
	\caption{Types and initialization of \PBqueue.}
	\label{alg:pbqueue-types}
\end{figure}

\begin{algorithm}
	\removelatexerror
	\scriptsize
	\begin{flushleft}
	\end{flushleft}	
	\setcounter{AlgoLine}{0}
	
	
	\myproc{ReturnValue {\footnotesize \PBqueue}(Function $func$, Data $arg$, Integer $seq$)}{
\nl		\lIf {$func = \Enqueue$} {\PBqueueEnq($args$, $seq$)}
\nl		\lElse {\PBqueueDeq($seq$)}
	}
	

	\myproc{{\footnotesize \PBqueueEnq}(Data $arg$, Integer $seq$)}{
		\tcp{Announce request}
		$ERequest[p] := \langle \Enqueue, Args, seq, 1 - ERequest[p].activate,1 \rangle$\;			\label{alg:pbqueue-full-enq:announce}
		\KwRet\ \PerformEnqueueRequest()
	}

	
	\myproc{ReturnValue {\footnotesize \PerformEnqueueRequest}()}{
		Bit $ind$   \tcp*{local variable of thread $p$} 
		Integer $lval$ \tcp*{local variable of thread $p$}
		\While{\True} {																					\label{alg:pbqueue-full-enq:while}
			$lval := ELock$ \;
			\uIf {$lval$ \Mod $2 = 0$} {								
				\lIf {$CAS(ELock, lval, lval+1) = \True$} { \Break }
				$lval := lval+1$
			}
			\WaitUntil $ELock \neq lval$ \;																\label{alg:pbqueue-full-enq:busy_wait}
			\uIf {$ERequest[p].activate = EState[EIndex].Deactivate[p]$} {						\label{alg:pbqueue-full-enq:if_req_processed}
				\lIf {$ELockVal \neq lval$} {
					\WaitUntil $ELock \neq lval+2$
				}
				\KwRet $EState[EIndex].ReturnVal[p]$;												\label{alg:pbqueue-full-enq:return}
			}
		}
	
		$ind: = 1 - EIndex$ \;																		\label{alg:pbqueue-full-enq:combiner:flip_index}
		$EState[ind] := EState[EIndex]$          \tcp*{create a copy of current state}			\label{alg:pbqueue-full-enq:combiner:state_copy}

		\For{$q\gets0$ \KwTo $n-1$} {																	\label{alg:pbqueue-full-enq:combiner:for:start}
			\tcp{if $q$ has a request that is not yet Applied}
			\uIf {$ERequest[q].valid=1$ \And\ $ERequest[q].activate \neq EState[ind].Deactivate[q]$} {        					\label{alg:pbqueue-full-enq:combiner:if_apply}
				\lred{\Add $EState[ind].Tail$ \To $toPersist$} \;							\label{alg:pbqueue-full-enq:combiner:defer_persist_LL_modifications_1}
				\Enqueue($\&EState[ind].Tail$, $ERequest[q].arg$)							\label{alg:pbqueue-full-enq:apply} \;
				$EState[ind].ReturnVal[q] := ACK$ \;
				$EState[ind].Deactivate[q] := ERequest[q].activate$
			}
		}
		\lred{\Add $EState[ind].Tail$ \To $toPersist$} \;							\label{alg:pbqueue-full-enq:combiner:defer_persist_LL_modifications_2}
		\lred{\lForEach{$e \in toPersist$}{\pwb{$e$}}}								\label{alg:pbqueue-full-enq:combiner:persist_LL_modifications}
		\pwb{$\&EState[ind]$} \;																		\label{alg:pbqueue-full-enq:persist_EState}
		\pfence{} \;  																			\label{alg:pbqueue-full-enq:sync_EState}	
		$ELockVal:=ELock$\;
		$EIndex := ind$ \;																			\label{alg:pbqueue-full-enq:combiner:updateS}
		\pwb{$\&EIndex$}																				\label{alg:pbqueue-full-enq:persist_EIndex}  \;		
		\psync{}																						\label{alg:pbqueue-full-enq:sync_EIndex} \;
		\lred{$oldTail := EState[ind].Tail$} \;													\label{alg:pbqueue-full-enq:combiner:oldTail_set}
		\lred{$toPersist = \emptyset$ \;}																\label{alg:pbqueue-full-enq:combiner:toPersist_reset}
		$ELock := ELock + 1$ \;																			\label{alg:pbqueue-full-enq:combiner:unlock}
		\KwRet $EState[EIndex].ReturnVal[p]$														\label{alg:pbqueue-full-enq:combiner:return}
	}


	\myproc{{\footnotesize \Enqueue}(Node **$Tail$, Data $arg$)}{
\nl		Node *$newnode :=$ \Allocate a new structure Node \;
\nl		$newnode \rightarrow data := arg$ \;
\nl		$newnode \rightarrow next := \bot$ \;
\nl		$($ *$Tail) \rightarrow next := newnode$\; 
\nl		*$Tail := newnode$
	}

	\caption{\PBqueue\ -- Code of \PBqueue\ and \PBqueueEnq\ for thread $p \in \{0,\ldots, n-1\}$}
	\label{alg:pbqueue-full-funcs-enq}
\end{algorithm}

\begin{algorithm}
	\removelatexerror
	\scriptsize
	\begin{flushleft}
	\end{flushleft}	
	

	\myproc{Node *{\footnotesize \PBqueueDeq}(Integer $seq$)}{
		\tcp{Announce request}
		$DRequest[p] := \langle \Dequeue, Args, seq, 1 - DRequest[p].activate,1 \rangle \rangle$\;	            \label{alg:pbqueue-full-deq:announce}
		\KwRet\ \PerformRequest()
	}

	
	\myproc{ReturnValue {\footnotesize \PerformRequest}()}{
\nl		Bit $ind$   \tcp*{local variables of thread $p$} 
		Integer $lval$ \tcp*{local variable of thread $p$}
		\While{\True} {																					\label{alg:pbqueue-full-deq:while}
			$lval := DLock$ \;
			\uIf {$lval$ \Mod $2 = 0$} {								
				\lIf {$CAS(DLock, lval, lval+1) = \True$} { \Break }
				$lval := lval+1$
			}
			\WaitUntil $DLock \neq lval$ \;																\label{alg:pbqueue-full-deq:busy_wait}
			\uIf {$DRequest[p].activate = DState[DIndex].Deactivate[p]$} {						\label{alg:pbqueue-full-deq:if_req_processed}
				\lIf {$DLockVal \neq lval$} {
					\WaitUntil $DLock \neq lval+2$
				}
				\KwRet $DState[DIndex].ReturnVal[p]$;												\label{alg:pbqueue-full-deq:return}
			}
		}
	
		$ind: = 1 - DIndex$ \;																		\label{alg:pbqueue-full-deq:combiner:flip_index}
		$DState[ind] := DState[DIndex]$          \tcp*{create a copy of current state}			\label{alg:pbqueue-full-deq:combiner:state_copy}

		\tcp{Simulate the dequeues}
		\For{$q\gets0$ \KwTo $n-1$} {																	\label{alg:pbqueue-full-deq:combiner:for:start}
			\tcp{if $q$ has a request that is not yet applied}
			\uIf {$DRequest[q].valid=1$ \And $DRequest[q].activate \neq DState[ind].Deactivate[q]$} {        					\label{alg:pbqueue-full-deq:combiner:if_apply}
				\lred{\uIf{$oldTail \neq DState[ind].Head$}{										\label{alg:pbqueue-full-enq:combiner:oldTail_not_reached}
					\lblack{$returnVal :=$ \Dequeue($\&DState[ind].Head$)} 	\label{alg:pbqueue-full-deq:apply}
				}
				\lElse {$returnVal := \bot$}}														\label{alg:pbqueue-full-enq:combiner:oldTail_reached}
				$DState[ind].ReturnVal[q] := returnVal$ \;
				$DState[ind].Deactivate[q] := DRequest[q].activate$
			}
		}

		\pwb{$\&DState[ind]$} \;																		\label{alg:pbqueue-full-deq:persist_DState}
		\pfence{} \;																\label{alg:pbqueue-full-deq:sync_DState}	
		$DLockVal:=DLock$\;
		$DIndex := ind$ \;																			\label{alg:pbqueue-full-deq:combiner:updateS}
		\pwb{$\&DIndex$} \;																			\label{alg:pbqueue-full-deq:persist_DIndex}	
		\psync{} \;																						\label{alg:pbqueue-full-deq:sync_DIndex}
	
		$DLock := DLock + 1$ \;																			\label{alg:pbqueue-full-deq:combiner:unlock}
		\KwRet $DState[DIndex].ReturnVal[p]$\;														\label{alg:pbqueue-full-deq:combiner:return}
	}


	\myproc{Node *{\footnotesize \Dequeue}(Node **$Head$)}{
\nl		Node *$ret := ($ *$Head) \rightarrow next$ \;
\nl		\lIf {$ret \neq \bot$} {*$Head := ($ *$Head) \rightarrow next$}
\nl		\KwRet $ret$
	}

	\caption{\PBqueue\ -- Code of \PBqueueDeq\ for thread $p \in \{0,\ldots, n-1\}$}
	\label{alg:pbqueue-full-funcs-deq}
\end{algorithm}

\begin{algorithm}
	\removelatexerror
	\scriptsize
	\begin{flushleft}
	\end{flushleft}			

	\myproc{ReturnValue {\footnotesize \Recover}(Function $func$, Data $arg$, Integer $seq$)}{
\nl		\lred{$lTail := EState[EIndex].Tail$}\;
\nl		\lred{$\CAS(\&oldTail, \&DUMMY, lTail)$}\;
\nl		\uIf{$func = \Enqueue$} {
\nl			$ERequest[p]:= \langle func, args, seq \mod 2, 1\rangle$\;
			\tcp{if request is not yet applied}
			\uIf {$EState[EIndex].Deactivate[p] \neq seq \mod 2$ } {								\label{alg:pbqueue-full-enq:recovery:rec-not-applied}
				\KwRet\ \PerformEnqueueRequest()													\label{alg:pbqueue-full-enq:recovery:re-execute-rec}
			}
			\tcp{request is applied}
			\KwRet $EState[EIndex].ReturnVal[p]$   													\label{alg:pbqueue-full-enq:recovery:rec-applied}		
		}
\nl		\Else{
\nl			$DRequest[p]:= \langle func, args, seq \mod 2, 1\rangle$\;
			\tcp{if request is not yet applied}
			\uIf {$DState[DIndex].Deactivate[p] \neq seq \mod 2$ } {								\label{alg:pbqueue-full-deq:recovery:rec-not-applied}
				\KwRet\ \PerformDequeueRequest()													\label{alg:pbqueue-full-deq:recovery:re-execute-rec}
			}
			\tcp{request is applied}
			\KwRet $DState[DIndex].ReturnVal[p]$   													\label{alg:pbqueue-full-deq:recovery:rec-applied}		
		}
	}
	\caption{\PBqueue\ -- Code of \Recover\ for thread $p \in \{0,\ldots, n-1\}$}
	\label{alg:pbqueue-full-funcs-rec}
\end{algorithm}

A combiner in $I_E$ serves only enqueue requests, while a combiner
in $I_D$ serves only dequeue requests.
$I_E$ stores just the tail of the queue in the $st$ field of its $StateRec$ records,
whereas $I_D$ stores just the head.  
We denote by $EStateRec$ ($DStateRec$), $ERequest$ ($DRequest$), $EState$ ($DState$), 
and $EIndex$ ($DIndex$) the $StateRec$, $Request$, $State$, and $MIndex$ of \PBcomb\ used by $I_E$ ($I_D$),
as shown in Figure~\ref{alg:pbqueue-types}. 
Algorithms~\ref{alg:pbqueue-full-funcs-enq}-\ref{alg:pbqueue-full-funcs-deq} provide 
the codes for the enqueuers, the dequeuers, and the standard sequential codes of the \Enqueue\ 
and the \Dequeue\ operations.
The recovery function (which is similar to that of \PBcomb) in presented in Algorithm~\ref{alg:pbqueue-full-funcs-rec}.
Parts that differentiate from \PBcomb\  are highlighted in red 
and is responsible for persistence (e.g., it persists the nodes of $L$), as described below.
The rest of the code resembles that of \PBcomb\ and it is easy to follow. 

The persistence scheme of \PBcomb\ 
guarantees that the head and the tail 
of the simulated queue are persistent.
In order to persist the changes performed on $L$, each enqueue combiner maintains a set 
$toPersist$ containing pointers to those nodes of $L$
that have either been updated (specifically, its $next$ pointer) or been created 
by the combiner (lines~\ref{alg:pbqueue-full-enq:combiner:defer_persist_LL_modifications_1} 
and~\ref{alg:pbqueue-full-enq:combiner:defer_persist_LL_modifications_2}).
After its simulation phase, a combiner persists its
modifications on $L$ by executing a \pwbi\ for each element of $toPersist$
(line~\ref{alg:pbqueue-full-enq:combiner:persist_LL_modifications}).
These \pwbi s are necessary; otherwise, the modifications performed by the
combiner on $L$ will not survive after a crash, which may result in an inconsistent 
state and violate durable linearizability. 
(We allocate new nodes in contiguous memory regions, thus these \pwbi s persist
as few cache lines as possible.)

Note that a \Dequeue\ operation only updates the head of the simulated queue 
and does not modify the nodes of $L$. 
Therefore, the effects of a dequeue combiner on the simulated state of the queue are 
correctly persisted by \PBqueue.
However, there is the following subtlety that needs to be addressed, 
regarding the nodes of $L$ that can be removed by the dequeue combiners.
An enqueue combiner simulates the active \Enqueue\ requests by directly modifying 
the nodes of $L$ and persists these modifications later.
Thus, if no care is taken, a dequeue combiner $p_d$ may remove from $L$ nodes that have been 
appended by an active enqueue combiner $p_e$.
Then, the dequeue combiner may complete (and thus the corresponding served \Dequeue\ 
operations may respond) before the enqueue combiner persists its modifications. 
This jeopardizes detectability. Specifically, 
assume that $p_d$ has removed a node containing data $d$ while simulating a \Dequeue\ request $D$, 
which has been appended in the list by $p_e$ while simulating an \Enqueue($d$)\ request $E$.
The process that announced $D$ responds with $d$ before $p_e$ persists its modifications while simulating $E$.
If a crash occurs, during recovery, the simulated state of the 
queue will not contain the response of $E$ and thus $E$ is considered not applied, after the crash.
So, after the crash, it is like $E$ has never enqueued $d$, since otherwise, it should be able
to find its response. However, $D$ has already responded with $d$ before the crash.
Durable linearizability still holds but detectability is violated.

To address this, \PBqueue\ disallows dequeue combiners to remove any node of $L$ that has not yet been 
persisted. It achieves this by using a shared volatile variable $oldTail$,
which initially points to $DUMMY$. 
Just before releasing the lock, an enqueue combiner 
sets $oldTail$ to point to the last node of $L$ (line~\ref{alg:pbqueue-full-enq:combiner:oldTail_set}). 
Then, the dequeue combiner removes nodes from $L$ as long as $oldTail$ is not reached
(lines~\ref{alg:pbqueue-full-enq:combiner:oldTail_not_reached}-\ref{alg:pbqueue-full-deq:apply}).
When $oldTail$ is reached, the dequeue combiner reports that the queue is empty for the 
corresponding \Dequeue\ requests (line \ref{alg:pbqueue-full-enq:combiner:oldTail_reached}).

\section{Wait-Free Recoverable Queue}
\label{appendix:PWFqueue}
\PWFqueue\ borrows
ideas from \SimQueue~\cite{FK11spaa,FK14}, so the description 
below follows that in~\cite{FK11spaa,FK14}. 
\PWFqueue\ uses two instances of \PWFcomb. The first is used to synchronize
the enqueuers and the second to synchronize the dequeuers. To achieve
independent execution of the enqueuers from the dequeuers, the first node in the list
is always considered as a dummy node.
We denote by $EStateRec$, $ERequest$, $EState$, $ES$, $DFlush$, and $DCombRound$ the instances of $StateRec$,
$Request$, $State$, $S$, $Flush$, and $CombRound$ of \PWFcomb\
used by the enqueuers. Similarly,
we denote by $DStateRec$, $DRequest$, $DState$, $DS$, $DFlush$, and $DCombRound$ the instances of $StateRec$,
$Request$, $State$, $S$, $Flush$, and $CombRound$ of \PWFcomb\ used by the dequeuers.
The queue is implemented by a singly-linked list, 
where $DS \rightarrow Head$ points to the head node of the queue
and $ES \rightarrow Tail$ points to the tail node of the queue. 

\begin{figure}[t]
	\scriptsize		
	
	\begin{flushleft}

		type RequestRec \{ \\			\label{alg:pwfqueue:AnnounceRec-start}
		\hspace*{6mm} $\{\Enqueue, \Dequeue\}$ $func$ \\
		\hspace*{6mm} Integer $arg$ \\
		\hspace*{6mm} Bit $activate$ \\
		\hspace*{6mm} Bit $valid$\\
		\}							\label{alg:pwfqueue-full:AnnounceRec-end}		

		type Node \{ \\
			\hspace*{6mm} Data $data$ \\
			\hspace*{6mm} Node *$next$ \\			
		\}

		type EStateRec \{ \\
		\hspace*{6mm} Node *$Tail$ \\
		\hspace*{6mm} \lred{Node *$oldTail$} \\		
		\hspace*{6mm} \lred{Node *$lhead$} \\
		\hspace*{6mm} ReturnValue $ReturnVal[0..n-1]$ \\
		\hspace*{6mm} Bit $Deactivate[0..n-1]$ \\
		\hspace*{6mm} Bit $Index[0..n-1]$    \\
		\hspace*{6mm} $\{0,1,..,n-1\}$ $pid$ \\
		\}

		type DStateRec \{ \\
		\hspace*{6mm} Node *$Head$  \\
		\hspace*{6mm} ReturnValue $ReturnVal[0..n-1]$ \\
		\hspace*{6mm} Bit $Deactivate[0..n-1]$ \\
		\hspace*{6mm} Bit $Index[0..n-1]$    \\
		\hspace*{6mm} $\{0,1,..,n-1\}$ $pid$ \\
		\}

		\vspace*{1mm}
		\com Local volatile variable: \\
		\lred{Set$\langle$Node*$\rangle$ $newItems$} \\

		\vspace*{1mm}
		\com Shared non-volatile variable: \\
		Node $DUMMY$, initially $\langle \bot, \bot \rangle$ \\
		
		\vspace*{1mm}
		\com Shared non-volatile variables used by the \PWFenqueue\ instance of \PWFcomb: \\
		EStateRec $EState[0..n][0..1]$, initially $\langle \&DUMMY, \bot, \bot, \lblue{\bot}, \langle \bot,\ldots,\bot\rangle, \langle 0,\ldots,0\rangle, \langle 0,\ldots,0\rangle, 0\rangle$ \\		
		EStateRec *$ES := \&EState[n][0]$, initially $ES$ points to $EState[n][0]$ \\		

		\vspace*{1mm}
		\com Shared volatile variables used by the \PWFenqueue\ instance of \PWFcomb: \\
		RequestRec $ERequest[0..n-1]$, initially $\langle\langle\bot,\bot,0,0\rangle, \ldots, \langle\bot,\bot,0,0\rangle\rangle$ \\
		Integer $EFlush[0..n-1]$, initially $\langle 0, \ldots, 0\rangle$ \\
		Integer $ECombRound[0..n-1][0..n-1]$, initially $\langle\langle 0, \ldots, 0\rangle, \ldots, \langle 0, \ldots, 0\rangle\rangle$ \\

		\vspace*{1mm}
		\com Shared non-volatile variables used by the \PWFdequeue\ instance of \PWFcomb: \\
		DStateRec $DState[0..n][0..1]$, initially $\langle \&DUMMY, \langle \bot,\ldots,\bot\rangle, \langle 0,\ldots,0\rangle, \langle 0,\ldots,0\rangle, 0\rangle$ \\		
		DStateRec *$DS := \&DState[n][0]$, initially $DS$ points to $DState[n][0]$ \\		

		\vspace*{1mm}
		\com Shared volatile variables used by the \PWFenqueue\ instance of \PWFcomb: \\
		RequestRec $DRequest[0..n-1]$, initially $\langle\langle\bot,\bot,0,0\rangle, \ldots, \langle\bot,\bot,0,0\rangle\rangle$ \\
		Integer $DFlush[0..n-1]$, initially $\langle 0, \ldots, 0\rangle$ \\
		Integer $DCombRound[0..n-1][0..n-1]$, initially $\langle\langle 0, \ldots, 0\rangle, \ldots, \langle 0, \ldots, 0\rangle\rangle$ \\

	\end{flushleft}	
	\caption{Types and initialization of \PWFqueue.}
	\label{alg:pwfqueue-full-types}
\end{figure}

\begin{algorithm}[t]
	\removelatexerror
	\scriptsize
	\begin{flushleft}
	\end{flushleft}	
	\setcounter{AlgoLine}{0}
	
	
		\begin{procedure}[H]
		\caption{() ReturnValue \PWFqueue(Function $func$, Data $arg$, Integer $seq$)}
		\lIf {$func = \Enqueue$} {\PWFenqueue($arg$, $seq$)}
		\lElse {\PWFdequeue($seq$)}
	\end{procedure}
	

	\begin{procedure}[H]
		\caption{() \PWFenqueue(Data $arg$, Integer $seq$)}
		\tcp{Announce request}
		$ERequest[p] := \langle \Enqueue, arg, seq, 1 - ERequest[p].activate, 1 \rangle$ \;	    \label{alg:pwfqueue-full-enq:announce}
		Backoff() \;						                										\label{alg:pwfqueue-full-enq:backoff}
		\KwRet \PerformEnqueueRequest()
	\end{procedure}


	\begin{procedure}[H]
		\caption{() Node *\PWFdequeue(Integer $seq$)}
		\tcp{Announce request}
		$DRequest[p] := \langle \Dequeue, \bot, seq, 1 - DRequest[p].activate, 1 \rangle$ \;	    \label{alg:pwfqueue-full-deq:announce}
		Backoff() \;						                									\label{alg:pwfqueue-full-deq:backoff}
		\KwRet \PerformDequeueRequest()
	\end{procedure}

	
	\begin{procedure}[H]
		\caption{() ReturnValue \Recover(Function $func$, Data $arg$, Integer $seq$)}
		\uIf{$func = \Enqueue$} {
			$ERequest[p]:= \langle func, args, seq \mod 2, 1\rangle$\;
			\tcp{if request is not yet applied}
			\uIf {$ES \rightarrow Deactivate[p] \neq seq \mod 2$} {			\label{alg:pwfqueue-full-enq:recovery:rec-not-Applied}
				Backoff() \;
				\KwRet\ \PerformEnqueueRequest()											\label{alg:pwfqueue-full-enq:recovery:PerformRequest}	
			}	
			\tcp{request is applied}
			\KwRet $ES \rightarrow ReturnVal[p].ret$  									\label{alg:pwfqueue-full-enq:recovery:rec-applied}		
		}
		\Else{
			$DRequest[p]:= \langle func, args, seq \mod 2, 1\rangle$\;
			\tcp{if request is not yet applied}
			\uIf {$DS \rightarrow Deactivate[p] \neq seq \mod 2$} {			\label{alg:pwfqueue-full-deq:recovery:rec-not-Applied}
				Backoff() \;
				\KwRet\ \PerformDequeueRequest()														\label{alg:pwfqueue-full-deq:recovery:PerformRequest}	
			}	
			\tcp{request is applied}
			\KwRet $DS \rightarrow ReturnVal[p].ret$  									\label{alg:pwfqueue-full-deq:recovery:rec-applied}	
		}
	\end{procedure}
	\caption{\PWFqueue\ -- Code of \PWFqueue, \PWFenqueue, \PWFdequeue, and \Recover\ for process $p \in \{0,\ldots, n-1\}$}
	\label{alg:pwfqueue-full-funcs}
\end{algorithm}	

\begin{algorithm}[t]
	\removelatexerror
	\scriptsize
	\begin{flushleft}
	\end{flushleft}	

	
	\begin{procedure}[H]
		\caption{() ReturnValue \PerformEnqueueRequest()}	
		EStateRec *$lsPtr$ \;	
		Integer $lval$\;
		\For{$l\gets1$ \KwTo $2$} {                                                 				\label{alg:pwfqueue-full-enq:two_attempts}
			$lsPtr := \LL(ES)$ \;                   			\label{alg:pwfqueue-full-enq:ll}
			Bit $ind := lsPtr\rightarrow Index[p]$ \;															\label{alg:pwfqueue-full-enq:index-read}
			$EState[i][ind] := $ *$lsPtr$      \tcp*{copy current state} 							\label{alg:pwfqueue-full-enq:copy_state}
			$EState[p][ind].pid := p$\;
			$lval := EFlush[lsPtr \rightarrow pid]$\;			\label{alg:pwfqueue-full-enq:Flush:read}
			\lIf {$lval \mod 2 = 0$} {
				$lval := lval +1$
			}
			\lElse {
				$lval := lval+2$								\label{alg:pwfqueue-full-enq:lval:init}
			}								

			\lIf {$\VL(ES) = \False$} {\Continue}                                    				\label{alg:pwfqueue-full-enq:vl}

			\lred{\uIf {$EState[p][ind].oldTail \neq \bot$ \And $EState[p][ind].oldTail \rightarrow next = \bot$} {
				\EnqueueConnect($\&EState[p][ind]$)													\label{alg:pwfqueue-full-enq:connect}
			}}
			\For{$q\gets0$ \KwTo $n-1$} {															\label{alg:pwfqueue-full-enq:apply_all}
				\tcp{if $q$ has a request that is not yet Applied}
				\uIf {$ERequest[q].valid=1$ \And $ERequest[q].activate \neq EState[p][ind].Deactivate[q]$} {       		\label{alg:pwfqueue-full-enq:if_apply}
					\Enqueue($\&EState[p][ind], ERequest[q].arg$)\;						\label{alg:pwfqueue-full-enq:apply}
					$EState[p][ind].ReturnVal[q] := ACK$ \;								\label{alg:pwfqueue-full-enq:response}
					$EState[p][ind].Deactivate[q] := ERequest[q].activate$\;						\label{alg:pwfqueue-full-enq:deactivate}
					$ECombRound[p][q] := lval$ \label{alg:pwfqueue-full-enq:set}					
				}
			}																						\label{alg:pwfqueue-full-enq:apply_all-end}

			\uIf {$\VL(ES) = \True$}{																\label{alg:pwfqueue-full-enq:vl-2}
				$EState[p][ind].Index[p] := 1 - EState[p][ind].Index[p]$ \; 					\label{alg:pwfqueue-full-enq:index-update}
				\lred{\lForEach{$e \in newItems$}{\pwb{$e$}}}								\label{alg:pwfqueue-full-enq:combiner:persist_LL_modifications}
				\pwb{\&$EState[p][ind]$} \;														\label{alg:pwfqueue-full-enq:EState:pwb}			
				\pfence{} \;																			\label{alg:pwfqueue-full-enq:EState:psync}
				$EFlush[p] := lval$\;
				\uIf {$\SC(ES, \&EState[p][ind]) = \True$ \tcp*{Try to change the contents of $ES$}}  {          							\label{alg:pwfqueue-full-enq:sc}
					\pwb{\&$ES$} \;																\label{alg:pwfqueue-full-enq:ES:pwb}
					\psync{} \;																		\label{alg:pwfqueue-full-enq:ES:psync}
					$\CAS(\&EFlush[p], lval, lval+1)$ \;		\label{alg:pwfqueue-full-enq:flush:reset}
					\KwRet $ES \rightarrow ReturnVal[p]$				
				}
				BackoffCalculate();
			}
		}
		
		$lsPtr := ES$ \;
		$lval := EFlush[lsPtr \rightarrow pid]$\;					\label{alg:pwfqueue-full-enq:EFlush:read-2}
		\uIf {$lval \mod 2 = 1$ \And $lval = ECombRound[lsptr\rightarrow pid][p]$} {\label{alg:pwfqueue-full-enq:flush:helper:read}
			\pwb{\&$ES$} \;								\label{alg:pwfqueue-full-enq:ES:pwb-2}
			\psync{} \;								\label{alg:pwfqueue-full-enq:ES:psync-2}
			$\CAS(EFlush[p], lval, lval+1)$\label{alg:pwfqueue-full-enq:flush:helper:reset}
		}
		\KwRet $ES \rightarrow ReturnVal[p]$						\label{alg:pwfqueue-full-enq:end}
	\end{procedure}
	

	\begin{procedure}[H]
		\caption{() \Enqueue(EStateRec *$lS$, Data $arg$)}
		Node *$newnode :=$ \Allocate a new structure Node \;
		$newnode \rightarrow data := arg$ \;
		$newnode \rightarrow next := \bot$ \;
		\lred{\Add $newnode$ \To $newItems$ \;
		\uIf{$lS \rightarrow oldTail = \bot$} {
			$lS \rightarrow oldTail := lS \rightarrow Tail$ \;
			$lS \rightarrow oldTailNext := newnode$ \;
		}}
		$lS \rightarrow Tail \rightarrow next := newnode$ \;
		$lS \rightarrow Tail := newnode$
	\end{procedure}

	
	\begin{procedure}[H]
		\caption{() \lred{\EnqueueConnect(EStateRec *$lS$)}}
		\lred{\uIf {$lS \rightarrow oldTail \rightarrow next = \bot$} {
			$\CAS(\&(lS \rightarrow oldTail \rightarrow next), \bot, \Mark(lS \rightarrow lhead))$
		}
		\uIf {\IsMarked($lS \rightarrow oldTail \rightarrow next$) = \True} {
			\pwb{$\&(lS \rightarrow oldTail \rightarrow next)$} \;
			$lS \rightarrow oldTail \rightarrow next := \UnMark(lS \rightarrow oldTail \rightarrow next)$ \tcp*{only for performance}
		}}
	\end{procedure}

	\caption{\PWFqueue\ -- Code of \PerformEnqueueRequest, \Enqueue, and \EnqueueConnect\  for process $p \in \{0,\ldots, n-1\}$}
	\label{alg:pwfqueue-full-funcs-enq}
\end{algorithm}

\begin{algorithm}[t]
	\removelatexerror
	\scriptsize
	\begin{flushleft}
	\end{flushleft}

	
	\begin{procedure}[H]
		\caption{() ReturnValue \PerformDequeueRequest()}	
		DStateRec *$lsPtr$ \;	
		Integer $lval$\;
		\For{$l\gets1$ \KwTo $2$} {                                                 			\label{alg:pwfqueue-full-deq:two_attempts}
			$lsPtr := \LL(DS)$ \;                			\label{alg:pwfqueue-full-deq:ll}
			Bit $ind := lsPtr\rightarrow Index[p]$ \;														\label{alg:pwfqueue-full-deq:index-read}
			$DState[i][ind] := $ *$lsPtr$      \tcp*{copy current state} 						\label{alg:pwfqueue-full-deq:copy_state}
			$DState[p][ind].pid := p$\;
			$lval := DFlush[lsPtr \rightarrow pid]$\;			\label{alg:pwfqueue-full-deq:Flush:read}
			\lIf {$lval \mod 2 = 0$} {
				$lval := lval +1$
			}
			\lElse {
				$lval := lval+2$								\label{alg:pwfqueue-full-deq:lval:init}
			}								
			\lIf {$\VL(DS) = \False$} {\Continue}                                    			\label{alg:pwfqueue-full-deq:vl}

			\For{$q\gets0$ \KwTo $n-1$} {														\label{alg:pwfqueue-full-deq:apply_all}
				\tcp{if $q$ has a request that is not yet Applied}
				\uIf {$DRequest[q].valid=1$ \And $DRequest[q].activate \neq DState[p][ind].Deactivate[q]$} {        	\label{alg:pwfqueue-full-deq:if_apply}
					\lred{\uIf {$DState[p][ind].Head \rightarrow next = \bot$} { $\DequeueConnect()$ }		\label{alg:pwfqueue-full-deq:connect}
					\uIf{$DState[p][ind].Head \rightarrow next \neq \bot$} {
						\lblack{$returnVal :=$ \Dequeue($\&DState[p][ind].Head$)} 						\label{alg:pwfqueue-full-deq:apply} \;
					}
					\lElse {$returnVal := \bot$}}
					$DState[p][ind].ReturnVal[q] := returnVal$ \;								\label{alg:pwfqueue-full-deq:response}
					$DState[p][ind].Deactivate[q] := DRequest[q].activate$
					$DCombRound[p][q] := lval$ \label{alg:pwfqueue-full-deq:set}					
				}
			}																					\label{alg:pwfqueue-full-deq:apply_all-end}
		
			\uIf {$\VL(DS) = \True$}{															\label{alg:pwfqueue-full-deq:vl-2}
				$DState[p][ind].Index[p] := 1 - DState[p][ind].Index[p]$ \; 				\label{alg:pwfqueue-full-deq:index-update}
				\pwb{\&$DState[p][ind]$} \;													\label{alg:pwfqueue-full-deq:DState:pwb}			
				\pfence{} \;																		\label{alg:pwfqueue-full-deq:DState:psync}
				$DFlush[p] := lval$\;
				\tcp{Try to change the contents of $DS$}
				\uIf {$\SC(DS, \&DState[p][ind]) = \True$} {          							\label{alg:pwfqueue-full-deq:sc}
					\pwb{\&$DS$} \;																\label{alg:pwfqueue-full-deq:DS:pwb}
					\psync{} \;																	\label{alg:pwfqueue-full-deq:DS:psync}
					$\CAS(\&DFlush[p], lval, lval+1)$ \;		\label{alg:pwfqueue-full-deq:flush:reset}
					\KwRet $DS \rightarrow ReturnVal[p]$				
				}
				BackoffCalculate();
			}
		}
		
		$lsPtr := DS$ \;
		$lval := DFlush[lsPtr \rightarrow pid]$\;					\label{alg:pwfqueue-full-deq:DFlush:read-2}
		\uIf {$lval \mod 2 = 1$ \And $lval = DCombRound[lsptr\rightarrow pid][p]$} {\label{alg:pwfqueue-full-deq:flush:helper:read}
			\pwb{\&$DS$} \;								\label{alg:pwfqueue-full-deq:DS:pwb-2}
			\psync{} \;								\label{alg:pwfqueue-full-deq:DS:psync-2}
			$\CAS(DFlush[p], lval, lval+1)$\label{alg:pwfqueue-full-deq:flush:helper:reset}
		}
		
		\KwRet $DS \rightarrow ReturnVal[p]$						\label{alg:pwfqueue-full-deq:end}
	\end{procedure}


	\begin{procedure}[H]
		\caption{() Node *\Dequeue(Node **$Head$)}
		Node *$ret := ($ *$Head) \rightarrow next$ \;
			\uIf {\IsMarked($ret$) = \True} {
				$ret := \UnMark(ret)$ \;
			}
			*$Head := ret$ \;
		\KwRet $ret$
	\end{procedure}

	
	\begin{procedure}[H]
		\caption{() \lred{\DequeueConnect()}}
		\lred{EStateRec *$lsPtr$ \;
		Node *$oldTail$ \;
		Node *$oldTailNext$ \;		
		
		$lsPtr := \LL(ES)$ \;
		$oldTail := lsPtr \rightarrow oldTail$ \;
		$oldTailNext := lsPtr \rightarrow oldTailNext$ \;		

		\uIf {$\VL(ES)$ \And $oldTail \rightarrow next = \bot$} {
			$\CAS(\&(oldTail \rightarrow next), \bot, \Mark(oldTailNext))$
		}}
	\end{procedure}
	
	\caption{\PWFqueue\ -- Code of \PerformDequeueRequest, \Dequeue, and \DequeueConnect\  for process $p \in \{0,\ldots, n-1\}$}
	\label{alg:pwfqueue-full-funcs-deq}
\end{algorithm}

Algorithm~\ref{alg:pwfqueue-full-funcs-enq} provides the code for the enqueuers, where parts that differentiate 
from \PWFcomb\ are identified in red.
A thread $p$ executing an enqueue request
will also try to serve enqueue requests by other enqueuers. It does so by creating
a local list of new nodes that will eventually be appended to the current state of the 
queue. Note that $EStateRec$ contains not only a pointer, $oldTail$, to the element reached
by following $next$ pointers starting from the head of the queue, but also 
two pointers, called $lHead$ and $Tail$, pointing to the first and last elements, respectively,
of the local list created by the last combiner that updated $ES$. 
We remark that at every point in time, the state of the queue
is comprised of all nodes (other than the dummy) that can be reached
by following next pointers starting from the node pointed to by $DS \rightarrow Head$, 
as well as of all nodes that can be reached by following the $next$ 
pointers starting from the node pointed to by $ES \rightarrow lHead$. Thus, 
$ES \rightarrow oldTail$ points to the node that was the tail of the queue before
the last \SC\ on $ES$ was performed, $ES \rightarrow Tail$ points to 
the current tail node of the queue, and $lHead$ points to a
node that will end up to be the node pointed to by $ES \rightarrow oldTail \rightarrow next$. 
The linking of $ES \rightarrow oldTail \rightarrow next$ to point to $ES \rightarrow lHead$
is performed by all enqueuers, by calling \EnqueueConnect\ (line~\ref{alg:pwfqueue-full-enq:connect}), 
before they proceed to serve any request.  

Algorithm~\ref{alg:pwfqueue-full-funcs-deq} provides the code for the dequeuers. The only subttlety 
is that a dequeuer should also ensure that the entire queue (which is now comprised
of two lists) is connected before it proceeds to serve the dequeues, as otherwise
consistency may be jeopardized. This is done by calling \DequeueConnect\ (line~\ref{alg:pwfqueue-full-deq:connect}). 
The rest of the code resembles that of \PWFcomb\ and is therefore easy to follow. 

The persistence scheme of \PWFqueue\ follows that of \PWFcomb. 
The persistence of \PWFcomb\ guarantees that each of its two instances 
utilized in \PWFqueue\ is persistent. The only subtlety comes from 
the necessity to connect the two parts of the linked list representing
the queue. 
An enqueuer has to persist the new value of the $next$ field
of the node it updates in \EnqueueConnect, for the following reason. 
Assume that two enqueue operations are executed by the same thread
without any dequeuer to take steps in the mean-time. Thus, new nodes 
have been connected to the linked list implementing the queue twice. 
If a crash occurs, upon recovery, the three pointers stored in $ES$ 
allows us to recover just the last of these connections (since 
$ES$ is persisted on lines~\ref{alg:pwfqueue-full-enq:ES:pwb}
and~\ref{alg:pwfqueue-full-enq:ES:pwb-2}).
This results in an inconsistent state. 

Note that although dequeuers also help connecting the two parts of the list,
they do not have to persist the $next$ field of the node $nd$ updated
in \DequeueConnect. Assume that a \Dequeue\ operation
has been persisted after traversing a link which has not been persisted yet. 
If the system fails, upon recovery, the head of the list has been 
persisted to point to a subsequent node to $nd$ 
and thus $nd$ will be never be accessed by any dequeuer again. 

\end{document}